\documentclass[11pt, reqno]{amsart} 
\usepackage{amssymb, amsmath}
\usepackage{hyperref}
\usepackage{enumerate}

\newcommand{\R}{\mathbb{R}}

\theoremstyle{plain}
\newtheorem{theorem}{Theorem}
\newtheorem{lemma}[theorem]{Lemma}
\newtheorem{prop}[theorem]{Proposition}
\newtheorem{cor}[theorem]{Corollary}

\theoremstyle{definition}

\newtheorem{definition}[theorem]{Definition}

\numberwithin{equation}{section}

\def\be{\begin{equation}}
\def\ee{\end{equation}}
\def\bea{\begin{eqnarray}}
\def\eea{\end{eqnarray}}

\begin{document}
\title[]{Symmetries of extremal horizons}
\author{Alex Colling}
\address{Department of Applied Mathematics and Theoretical Physics\\ 
University of Cambridge\\ Wilberforce Road, Cambridge CB3 0WA, UK.}
\email{aec200@cam.ac.uk}

\begin{abstract}
We prove an intrinsic analogue of Hawking's rigidity theorem for extremal horizons in arbitrary dimensions: any compact cross-section of a rotating extremal horizon in a spacetime satisfying the null energy condition must admit a Killing vector field. It follows that an extension of the near-horizon geometry admits an enhanced isometry group containing $SO(2,1)$ or the 2D Poincaré group $\mathbb{R}^2 \rtimes SO(1,1)$. In the latter case, the associated Aretakis instability for a massless scalar field is shifted by one order in the derivatives of the field transverse to the horizon. We consider a broad class of examples including Einstein-Maxwell(-Chern-Simons) theory and Yang-Mills theory coupled to charged matter. In these examples we show that the symmetries are inherited by the matter fields.
\end{abstract}

\maketitle

\section{Introduction}
Stationary black hole solutions to the Einstein equations have been a central topic of study in General Relativity for many decades. A key result in this context is Hawking's rigidity theorem \cite{H72,HE73, C96, FRW98}, which under certain assumptions (including analyticity) establishes that the event horizon of the black hole must be a Killing horizon. Moreover, if the black hole is rotating, i.e. the stationary Killing field is not normal to the horizon, the spacetime must be axially symmetric. Since the original proof by Hawking the result was extended to higher dimensions \cite{HIW07,MI08} and, under a condition on the angular velocities of the horizon, to the extremal case \cite{HI09}. The theorem remains valid within a wide class of matter theories.

The rigidity theorem relates the global concept of an event horizon (whose definition requires information about the spacetime asymptotics) to the locally defined notion of a Killing horizon. This paves the way for a quasi-local approach to studying black hole horizons using only the geometry of Killing horizons. Such an approach is naturally formulated within the framework of isolated horizons \cite{ABC98, ABL02, LP02, LP05} and near-horizon geometries \cite{R02, KL13}. There is a fundamental difference between extremal and non-extremal horizons. In the extremal case the Einstein equations imply a set of constraints, which we refer to as the \textit{horizon equations}, involving only data intrinsic to a spatial cross-section of the horizon. By contrast, in the non-extremal case the Einstein equations restricted to the horizon involve information about the spacetime embedding of the horizon and do not impose constraints on the intrinsic data.

An intrinsic analogue of the ``rotating implies axisymmetric" theorem for extremal horizons in vacuum (allowing for a cosmological constant) was recently proven by Dunajski and Lucietti \cite{DL}. Their proof uses the horizon equations to derive a divergence identity which, assuming compactness of the cross-section, shows the existence of a Killing vector field on the horizon. The arguments were subsequently generalised to four-dimensional Einstein-Maxwell theory in \cite{CKL} (see also \cite{KL24}). The main purpose of this paper is to establish the intrinsic rigidity theorem for extremal horizons in a spacetime of arbitrary dimension and with arbitrary matter content, subject only to the null energy condition.

The induced data on an $n$-dimensional cross-section $M$ of an extremal horizon in an $(n+2)$-dimensional spacetime consists of a Riemannian metric $g$ and a 1-form $X$, as well as a symmetric $(0,2)$ tensor $T$ and a function $U$ induced by the matter content. These are constrained by the horizon equations \cite{KL13}
\begin{equation} \label{nhe}
    R_{ab} = \frac 12X_aX_b - \nabla_{(a}X_{b)} + T_{ab} - \frac 1n(g^{cd}T_{cd} + 2U)g_{ab}.
\end{equation}
Here $R_{ab}$ is the Ricci tensor of the Levi-Civita connection $\nabla$ of $g$. A solution is called \textit{rotating} if the 1-form $X$ is not exact (see the discussion at the end of Section \ref{ddhsec}). 

Out of the data $(M,g,X,T,U)$ one can construct a (Lorentzian) $(n+2)$-dimensional spacetime, the near-horizon geometry, in such a way that the Einstein equations for this spacetime are equivalent to (\ref{nhe}). We will consider solutions to (\ref{nhe}) for which this associated near-horizon geometry satisfies the null energy condition (NEC). It may be verified (Lemma \ref{lemec}) that, if any spacetime containing an extremal Killing horizon satisfies the NEC, then so does the associated near-horizon geometry. In this sense we are imposing the weakest possible requirement.

We are now in a position to state the intrinsic rigidity result, which is proven in Section \ref{rig}.  

\begin{theorem}\label{thm1}
    Let $(g,X,T,U)$ be a rotating solution to the horizon equations \eqref{nhe} on a compact manifold $M$ without boundary, such that the associated near-horizon geometry satisfies the null energy condition. Then $(M,g)$ admits a Killing vector field $K$ preserving the horizon data $(g,X,T,U)$. 
\end{theorem} 

\noindent \textbf{Remark.} An earlier version of this result additionally assumed a version of the dominant energy condition imposed on null vectors. I am grateful to an anonymous referee, whose detailed suggestions made it possible to remove this assumption. \\

The fact that $K$ preserves the horizon data ensures it extends to a Killing vector field in the near-horizon geometry. Following the method in \cite{KL24, DL}, we deduce (Proposition \ref{propA}) the existence of a function $A$ on $M$ which is constant, regardless of whether the horizon is rotating or not. We proceed by showing that $A$ appears as the Gaussian curvature of a two-dimensional Lorentzian factor in the near-horizon geometry. This generalises the corresponding results in \cite{DL} for the vacuum case and in \cite{KLR} for spacetimes with isometry group $\mathbb{R}\times U(1)^{n-1}$ (see also \cite{L12}). Depending on the sign of $A$, we extend the 2D factor to the full two-dimensional anti-de Sitter (AdS$_2$), Minkowski ($\R^{1,1}$) or de Sitter (dS$_2$) space. The extended near-horizon geometry then admits an enhanced isometry group.
\begin{theorem} \label{thm2}
Any extended near-horizon geometry with compact cross-sections satisfying the null energy condition admits an isometry group containing the orientation-preserving isometries of \textup{AdS}$_2$, $\mathbb{R}^{1,1}$ or \textup{dS}$_2$. If the corresponding horizon data is rotating and the strong energy condition is satisfied, only the \textup{AdS}$_2$ case is possible.
\end{theorem}
For rotating horizons the isometry group has an additional $U(1)$ factor coming from the Killing vector in Theorem \ref{thm1}. A special case where the near-horizon geometry admits a further symmetry enhancement and locally has an AdS$_3$ factor is discussed in Section~\ref{ads3}.

The constant $A$ arises in the context of the Aretakis instability, which is an instability associated to the wave equation for a massless scalar field on an extremal horizon. It was originally identified for the extremal Reissner-Nordstr\"om spacetime in \cite{A11,A112}, where Aretakis showed that the first transverse derivative of the scalar field is conserved along the horizon and, for generic initial data, higher derivatives grow polynomially in the affine parameter $v$.  The proof uses a set of conservation laws which depend only on the local geometry of the horizon. The result was subsequently extended to arbitrary extremal horizons under the assumption that $A$ does not vanish \cite{LR12,A12}. There do however exist extremal horizons for which $A = 0$, and we argue in Section \ref{ddhsec} that these should be considered doubly degenerate. An example is given by the ``ultracold" Reissner-Nordstr\"om-de Sitter spacetime where the event horizon, Cauchy horizon and cosmological horizon coincide.

In Section~\ref{secare}, building on the method of \cite{LR12},  we show (Proposition \ref{areprop}) that the Aretakis instability is shifted by one order for a doubly degenerate horizon. Certain combinations of the scalar field and its first two transverse derivatives are conserved along the horizon, and, provided the field itself decays, a quantity involving a third derivative generically blows up as $v \to \infty$. This behaviour holds under a condition which may be interpreted as ensuring that the horizon is not triply degenerate. We verify this condition and compute the constant $A$ explicitly for the extremal Kerr-Newman-de Sitter family in Appendix~\ref{Aa}.

$A$ can be viewed as an extremal analogue of the surface gravity: it is constant as a consequence of the Einstein equations and vanishes for (doubly) degenerate configurations. In this sense Proposition ~\ref{propA} in Section ~\ref{senhsec} is an analogue of the zeroth law of black hole mechanics for extremal horizons. Further justification of this interpretation comes from the fact that $A$ plays the role of the surface gravity in near-horizon versions of the Smarr relation \cite{CL25}.

In Sections \ref{fsec} and \ref{gsec} we consider fairly general examples of matter models satisfying the NEC. These include many theories in which extremal horizons are of interest, such as supergravity theories and their dimensional reductions, Yang-Mills theory \cite{LL13} and Einstein-Maxwell theory coupled to charged matter, which is studied in the context of the third law of black hole mechanics \cite{KU22, R24}. We show how to use the field equations to prove that the horizon data induced by matter fields is preserved by the Killing vector constructed in Theorem~\ref{thm1}. Using this data it is possible to define matter fields in the near-horizon geometry, which are shown to be invariant under the isometries in Theorem~\ref{thm2}.

\subsection*{Acknowledgements} I would like to thank my PhD supervisor Maciej Dunajski and James Lucietti for many insightful discussions, as well as guidance and encouragement that helped shape this project. I am also grateful to Wojciech Kami\'nski, Christoph Kehle, Harvey Reall and Jun Liu for comments and discussions. Finally, I thank the anonymous referees for their comments, which led to a number of  improvements, including a strengthening of Theorem \ref{thm1}. I am supported by the Cambridge International Scholarship. This work has been partially supported by STFC consolidated grant ST/X000664/1.

\section{Preliminaries}
We consider solutions to the Einstein equations containing an extremal Killing horizon. The most relevant examples of such solutions are extremal black hole spacetimes, but, with the exception of Section \ref{ddhbigsec}, our analysis relies only on the intrinsic geometry of the horizon and therefore applies more generally to extremal isolated horizons as in \cite{LP02}. The following setup is based on \cite{KL13}, and details of computations can be found there. All objects are assumed to be smooth. Greek indices are used for the spacetime, while Latin indices refer to a cross-section of the horizon. Tensors with a subscript H are defined in the near-horizon geometry.

\subsection{Extremal horizons} \label{ehors}
Let $(\mathcal{M},\textbf{g})$ be a Lorentzian manifold of dimension $n+2$ satisfying the Einstein equations
\begin{equation} \label{Einstein}
    \mathcal{R}_{\mu\nu} - \frac 12(\textbf{g}^{\rho\sigma}\mathcal{R}_{\rho\sigma})\textbf{g}_{\mu\nu} =\mathcal{T}_{\mu\nu}.
\end{equation}
We allow for a cosmological constant, which we absorb in the effective energy-momentum tensor $\mathcal{T}_{\mu\nu}$. Suppose $(\mathcal{M},\textbf{g})$ contains an extremal Killing horizon $\mathcal{H}$. This means $\mathcal{H}$ is a null hypersurface and there exists a Killing vector $k$ of $(\mathcal{M},\textbf{g})$ which is normal (and also tangent) to $\mathcal{H}$. The vector $k$ is called the generator of $\mathcal{H}$. The extremality condition is that $\textbf{g}(k,k)$ has a double zero on $\mathcal{H}$, i.e.
\begin{equation}
    \textbf{g}(k,k) \:\overset{\mathcal{H}}{=}\:0, \hspace{1cm}\text{d}(\textbf{g}(k,k)) \:\overset{\mathcal{H}}{=} \:0.
\end{equation}
Here $\overset{\mathcal{H}}{=}$ denotes equality after evaluating on $\mathcal{H}$. Equivalently, the integral curves of $k$ are affinely parametrised null geodesics on $\mathcal{H}$. We assume $\mathcal{H}$ is diffeomorphic to $M \times \R$, where $M$ is an $n$-dimensional spacelike submanifold transversal to the integral curves of $k$. Importantly, $M$ is assumed to be compact and without boundary. Topologically non-trivial horizons that do not admit a global cross-section have been studied recently in \cite{BDLO25} (see also \cite{DLO23}). In our setting, we fix a choice of $M$, which then inherits the following data from $(\textbf{g},\mathcal{T})$.
\begin{enumerate}
    \item A Riemannian metric $g$ obtained by pulling back $\textbf{g}$ along the inclusion $i: M \to \mathcal{M}$.
    \item A 1-form $X$ defined as follows: since $k \wedge \text{d}k = 0$ on any Killing horizon, there exists a 1-form $\tilde{X}$ on $\mathcal{H}$ such that 
    \begin{equation}  \label{Xdef}
    \text{d}k \:\overset{\mathcal{H}}{=}\: k \wedge \tilde{X}.
\end{equation}
Note that we are using the same notation for $k$ and its $\textbf{g}$-dual 1-form. We have $\mathcal{L}_k\tilde{X} = 0$ and the extremality condition implies  $\iota_k\tilde{X} = 0$, so that we may write $\tilde{X} = \pi^* X$ for a 1-form $X$ on $M$, where $\pi: \mathcal{H} \to M$ denotes the natural projection.
\item A symmetric $(0,2)$ tensor $T$ obtained by pulling back the energy-momentum tensor $\mathcal{T}$ via $i$.
\item A function $U$ defined by
\begin{equation} \label{U}
    \mathcal{T}(k,\cdot)\:\overset{\mathcal{H}}{=}\: Uk.
\end{equation}
The Einstein equations imply that $U$ is well-defined, since inserting $k$ into the Ricci tensor gives a 1-form $\mathcal{R}(k,\cdot)$ that is proportional to $k$ on any extremal horizon\footnote{On any Killing horizon the 1-form $\mathcal{R}(k, \cdot)$ is proportional to $k$ if and only if the surface gravity $\kappa$ (defined by $^{\textbf{g}}\nabla_k k = \kappa \:k$ on $\mathcal{H}$) is constant on $\mathcal{H}$. For an extremal horizon $\kappa$ is identically zero.}.
\end{enumerate}
We refer to the data  $(g,X,T,U)$ as \textit{horizon data} on $M$. In a specific theory this may be supplemented by data induced by the matter fields in the theory (which are also assumed to be invariant under $k$). As a consequence of (\ref{Einstein}), the horizon data satisfies the horizon equations (\ref{nhe}).

Additional constraints on the matter data follow from the conservation of energy-momentum $\mathcal{W}_\mu = 0$, where $\mathcal{W}_\nu = {^\textbf{g}}\nabla^\mu\mathcal{T}_{\mu\nu}$ and ${^\textbf{g}}\nabla$ denotes the spacetime Levi-Civita connection. 
Consider the 1-form $\eta = \mathcal{T}(k,\cdot)-Uk$, which vanishes on $\mathcal{H}$ by the definition\footnote{In order to view $\eta$ as a 1-form on $\mathcal{M}$, we extend the function $U$ away from $M$ in any way such that $\mathcal{L}_k U = 0$.} (\ref{U}) of $U$. As $k$ is the normal to $\mathcal{H}$, any function vanishing on $\mathcal{H}$ must have exterior derivative proportional to $k$ on $\mathcal{H}$. Applying this argument to the coefficients of $\eta$ in any basis, it follows that there exists a 1-form $\tilde{\beta}$ on $\mathcal{H}$ so that
\begin{equation} \label{betadef}
\text{d}\eta \:\overset{\mathcal{H}}{=}\: k \wedge \tilde{\beta}.
\end{equation}
Since $\mathcal{R}(k,k)$ has a double zero on the horizon\footnote{It is well known that $\mathcal{R}(k,k)$ vanishes on any Killing horizon. In the extremal case this function actually has a double zero on $\mathcal{H}$, which is most easily seen in Gaussian null coordinates introduced below (see e.g. \cite{MI83} for the components of $\mathcal{R}$ in these coordinates).}, i.e. d$(\mathcal{R}(k,k)) = 0$ on $\mathcal{H}$, the Einstein equations (\ref{Einstein}) imply d$(\iota_k \eta) = \iota_k\text{d}\eta = 0$ on $\mathcal{H}$. Therefore $\iota_k\tilde{\beta} = 0$ and as before $\tilde{\beta} = \pi^* \beta$ for a well-defined 1-form $\beta$ on $M$. Writing $W = i^*\mathcal{W}$, we have the constraint
\begin{equation} \label{c1}
0 = W_a = \beta_a + \nabla^bT_{ab} + UX_a - X^bT_{ab}.
\end{equation}
Here indices are raised and lowered using the induced metric on $M$, and similarly the covariant derivative is taken with respect to $g$. From the fact that $\iota_k\eta$ has a double zero on the horizon we also deduce the existence of a function $\alpha$ such that 
\begin{equation} \label{alphadef}
    \text{Hess}_{\textbf{g}}(\iota_k\eta) \:\overset{\mathcal{H}}{=}\: 2\alpha \:k\otimes k.
\end{equation}
Furthermore, we have
\begin{equation*}
    \iota_k \mathcal{W} = {^{\textbf{g}}}\nabla^\nu (k^\mu\mathcal{T}_{\mu\nu}) = {^{\textbf{g}}}\nabla^\mu \eta_\mu \:\overset{\mathcal{H}}{=}\: 0.
\end{equation*}
It follows that there exists a function $\omega$ on $M$ such that $\text{d}(\iota_k\mathcal{W}) = 2\omega k$ holds on $\mathcal{H}$.
The second constraint coming from the conservation of $\mathcal{T}$ is
\begin{equation} \label{c2}
    0 = \omega = \alpha + \frac 12\nabla_a\beta^a - X^a\beta_a.
\end{equation}
In any specific theory the equations of motion for the induced matter fields on $M$ must imply (\ref{c1}) and (\ref{c2}). Finally, it is useful to introduce a function $F$ on $M$ by
\begin{equation}
    \text{Hess}_{\textbf{g}}(\textbf{g}(k,k)) \:\overset{\mathcal{H}}{=}\: 2F\:k\otimes k.
\end{equation}
This function describes the leading order behaviour of the norm of $k$ away from $\mathcal{H}$. The Einstein equations contracted once with $k$ evaluated on $\mathcal{H}$ allow us to express it in terms of horizon data as
\begin{equation}
    F = \frac 12\vert X \vert^2 - \frac 12 \nabla_a X^a + (1-\tfrac 2n)U - \frac 1n g^{ab}T_{ab}. \label{F}
\end{equation}
Here $\vert \cdot \vert$ denotes the $g$-norm. A solution $(g,X,T,U)$ to the horizon equations is called \textit{static} if d$X = 0$ and $\text{d}F = XF$, with $F$ as in (\ref{F}). These are the conditions for the near-horizon geometry defined in Section \ref{nhgs} to be static, i.e. for the generator $k$ to be hypersurface-orthogonal. The solution is called \textit{rotating} if $X$ is not exact. It is straightforward to verify with these definitions that, for example, the extremal Reissner-Nordstr\"om horizon is static and the extremal Kerr horizon is rotating. It is possible for a horizon to be both static and rotating, which will be discussed in Section \ref{ads3}.

\subsection{Near-horizon geometry} \label{nhgs} To any solution $(g,X,T,U)$ of the horizon equations we may associate a spacetime, the \textit{near-horizon geometry}, in the following way \cite{KL13}. We equip $\R^2 \times M$ with the metric and energy-momentum tensor
\begin{subequations} \label{nhg}
\begin{align} \label{nhga}
    \textbf{g}_{\text{H}} &= 2\text{d}v\text{d}r + 2rX\text{d}v + r^2 F\text{d}v^2 + g, \\
    \mathcal{T}_{\text{H}} &= 2U\text{d}v\text{d}r + 2r(\beta + UX)\text{d}v + r^2(\alpha + UF)\text{d}v^2 +T. \label{nhgb}
    \end{align}
\end{subequations}
Here $(v,r)$ are coordinates on $\R^2$. The data $(\beta,\alpha,F)$ is determined by the horizon data using (\ref{c1}, \ref{c2}, \ref{F}). This defines a spacetime containing an extremal Killing horizon $\mathcal{H} = \{r=0\}$ with generator $k = \partial_v$ whose horizon data recovers $(g,X,T,U)$. It may be verified that the Einstein equations for (\ref{nhg}) are equivalent to (\ref{nhe}). Note that in general a near-horizon geometry admits a two-dimensional isometry group generated by translations in $v$ and the scaling $(v,r)\mapsto (\lambda^ {-1}v,\lambda r)$, with corresponding Killing vectors $\partial_v$ and $v\partial_v - r\partial_r$.

It will be convenient to introduce a null-orthonormal frame
\begin{equation} \label{onbdef}
    e^+ = \text{d}v, \hspace{.8cm} e^- = \text{d}r + r X + \frac 12r^2F\text{d}v, \hspace{.8cm} e^i = \hat{e}^i.
\end{equation}
Here $\hat{e}^i$  $(1 \leq i \leq n)$ is an orthonormal basis for $g$ on $M$, i.e. $g = \delta_{ij}\:\hat{e}^i\hat{e}^j$. The dual basis is
\begin{equation} \label{onbdefvec}
    e_+ = \partial_v - \frac 12 F r^2 \partial_r, \hspace{.8cm} e_- = \partial_r, \hspace{.8cm} e_i = \hat{e}_i - r\hat{X}_i\partial_r,
\end{equation}
where $\hat{e}_i$ denotes the dual basis of $\hat{e}^i$ and $\hat{X}_i = \iota_{\hat{e}_i}X$. We can express (\ref{nhg}) as 
\begin{subequations} \label{onb}
\begin{align} \label{onba}
    \textbf{g}_{\text{H}} &= 2\:e^+ e^- + \delta_{ij}\:e^ie^j, \\
    \mathcal{T}_{\text{H}} &= 2U\:e^+ e^- + 2r\hat{\beta}_i\:e^i e^+ + r^2\alpha\: e^+ e^+ +\hat{T}_{ij}\:e^i e^j. \label{onbb}
    \end{align}
\end{subequations}
The near-horizon geometry may be obtained directly from the original spacetime $(\mathcal{M},\textbf{g},\mathcal{T})$ by a limiting procedure. Let us introduce Gaussian null coordinates $(v,r,x^i)$ in $\mathcal{M}$ around a point $p \in M$ such that the $x^i$ are local coordinates on $M$, the generator is $k = \partial_v$ and the horizon is at $r = 0$ (see \cite{MI83}). This defines a double foliation, and the $x^i$ extend to local coordinates on each leaf $M(v,r)$ of constant $(v,r)$. Gaussian null coordinates are uniquely determined by the choice of cross-section $M = M(0,0)$ and the coordinates $x^i$. The metric takes the form
\begin{equation} \label{ggnc}
    \textbf{g} = 2\text{d}v\text{d}r + 2rX_i(r,x)\text{d}x^i\text{d}v + r^2 F(r,x)\text{d}v^2 + g_{ij}(r,x)\text{d}x^i\text{d}x^j.
\end{equation}
Here $X_i, F$ and $g_{ij}$ are functions in a neighbourhood of $p$. The extremality condition corresponds to $\textbf{g}_{vv} = O(r^2)$. A null-orthonormal frame for $\textbf{g}$ is
\begin{equation} \label{onbs}
    \textbf{e}^+ = \text{d}v, \hspace{.8cm} \textbf{e}^- = \text{d}r + rX_i(r,x)\text{d}x^i(\hat{\textbf{e}}_j)\hat{\textbf{e}}^j + \frac 12r^2F(r,x)\text{d}v, \hspace{.8cm} \textbf{e}^i = \hat{\textbf{e}}^i,
\end{equation}
where $\hat{\textbf{e}}^i$ is an orthonormal basis on $M(v,r)$. For any $\epsilon > 0$, consider the transformation $\Psi_\epsilon$ given in the coordinate chart by $\Psi_\epsilon(v, r,x) = (\frac{v}{\epsilon},\epsilon r,x)$, and define the 1-parameter family of metrics
\begin{equation*}
    \textbf{g}_\epsilon = \Psi_\epsilon^*\textbf{g} = 2\text{d}v\text{d}r + 2rX_i(\epsilon r,x)\text{d}x^i\text{d}v + r^2 F(\epsilon r,x)\text{d}v^2 + g_{ij}(\epsilon r,x)\text{d}x^i\text{d}x^j.
\end{equation*}
Taking the limit $\epsilon \to 0$, we recover (\ref{nhga}) upon identifying $(v,r)$ with coordinates in the near-horizon geometry and setting 
\begin{equation} \label{ident}
    X = X_i(0,x)\text{d}x^i, \hspace{.8 cm} F = F(0,x), \hspace{.8cm} g = g_{ij}(0,x)\text{d}x^i\text{d}x^j.
\end{equation}
In other words, (\ref{nhga}) arises from (\ref{ggnc}) by evaluating the functions $X_i, F$ and $g_{ij}$ at $r = 0$.
It follows from the definitions in Section \ref{ehors} that the objects (\ref{ident}) do not depend on any choice of coordinates. We can similarly define $\mathcal{T}_\epsilon = \Psi_\epsilon^*\mathcal{T}$ and obtain $\mathcal{T}_{\text{H}} = \lim_{\epsilon \to 0}\mathcal{T}_\epsilon$ by considering the energy-momentum tensor in Gaussian null coordinates
\begin{equation}
    \mathcal{T} = 2U\text{d}v\text{d}r + 2r(\beta_i + UX_i)\text{d}x^i\text{d}v + r^2(\alpha+UF)\text{d}v^2 + T_{ij}\text{d}x^i\text{d}x^j + 2\mathcal{T}_{ri}\text{d}r\text{d}x^i + \mathcal{T}_{rr}\text{d}r\text{d}r.
\end{equation}
Here all components of $\mathcal{T}$ implicitly depend on $r$ and the $x^i$. The Einstein equations imply that $\mathcal{T}$ must be of this form (i.e. $\mathcal{T}_{vi} = O(r)$ and $\mathcal{T}_{vv} = O(r^2)$). We recover (\ref{nhgb}) in the limit by identifying 
\begin{equation}
    U = U(0,x), \hspace{.8cm} \beta = \beta_i(0,x)\text{d}x^i, \hspace{.8cm}\alpha = \alpha(0,x), \hspace{.8cm} T = T_{ij}(0,x)\text{d}x^i\text{d}x^j.
\end{equation}
Note that the components $\mathcal{T}_{rr}$ and $\mathcal{T}_{ri}$ do not contribute to the limit. If the energy-momentum tensor is determined by matter fields in the theory, there is no guarantee that the matter fields themselves have a well-defined near-horizon limit. We do not assume $\mathcal{T}_\text{H}$ is obtained from matter fields below, although we will see that this is the case for the theories considered in Sections \ref{fsec} and \ref{gsec}.

The near-horizon geometry can be thought of as a leading order approximation to the spacetime away from $\mathcal{H}$. In particular,  it inherits energy conditions like the NEC and the strong energy condition (SEC), which will be used later. Recall the strong energy condition is the requirement $\mathcal{R}(\ell,\ell) \geq 0$ for any timelike vector $\ell$, where $\mathcal{R}$ denotes the Ricci tensor. Equivalently,
\begin{equation}
(\mathcal{T}_{\mu\nu} - \tfrac 1n(\textbf{g}^{\rho\sigma}\mathcal{T}_{\rho\sigma})\textbf{g}_{\mu\nu})\ell^\mu\ell^\nu \geq 0.
\end{equation}
\begin{lemma} \label{lemec}
    If a spacetime $(\mathcal{M},\textbf{g},\mathcal{T})$ satisfies the NEC or the SEC, then so does the associated near-horizon geometry $(\R^2\times M, \textbf{g}_{\textup{H}},\mathcal{T}_{\textup{H}})$.
\end{lemma}
\begin{proof}
    Let $\ell$ be a vector in the near-horizon geometry at a point $p = (v_p,r_p,x_p)$, expressed in the basis (\ref{onbdefvec}) as 
    \begin{equation*}
        \ell = \ell^+e_++\ell^-e_- + \ell^ie_i.
    \end{equation*}
    Pushing $\ell$ forward by an isometry $(v,r,x) \mapsto (\lambda^{-1}v,\lambda r,x)$ of the near-horizon geometry if necessary, we may assume $r = r_p$ is in the range of the Gaussian null coordinate $r$ in $\mathcal{M}$. Define a vector $\ell^\epsilon$ in $\mathcal{M}$ by replacing the null-orthonormal frame of the near-horizon geometry by a null-orthonormal frame for $\textbf{g}_\epsilon$ at the point with Gaussian null coordinates $(v_p,r_p,x_p)$,
        \begin{equation*}
        \ell^\epsilon = \ell^+\textbf{e}_{+}^\epsilon+\ell^-\textbf{e}_-^\epsilon + \ell^i\textbf{e}_i^\epsilon.
    \end{equation*}
Explicitly,
\begin{equation*}
    \textbf{e}_+^\epsilon = \partial_v - \frac 12F(\epsilon r_p,x)r_p^2\partial_r, \hspace{.8cm} \textbf{e}_-^\epsilon = \partial_r, \hspace{.8cm} \textbf{e}_i^\epsilon = \hat{\textbf{e}}_i^\epsilon - r_pX_j(\epsilon r_p,x)\text{d}x^j(\hat{\textbf{e}}_i^\epsilon)\partial_r,
\end{equation*}
where the $\hat{\textbf{e}}^\epsilon_i$ are (dual to) a $\textbf{g}_\epsilon$-orthonormal frame for $M(v_p,r_p)$. The norm of $\ell$ with respect to  $\textbf{g}_{\text{H}}$ equals the norm of $\ell^\epsilon$ with respect to $\textbf{g}_\epsilon$. Moreover,
\begin{equation} \label{limeps}
    \lim_{\epsilon \to 0}      \mathcal{T}_\epsilon(\ell^\epsilon,\ell^\epsilon) = \mathcal{T}_{\text{H}}(\ell,\ell), \hspace{1cm} \lim_{\epsilon \to 0}    \textbf{g}_\epsilon^{\mu\nu}(\mathcal{T}_\epsilon)_{\mu\nu} = \textbf{g}_{\text{H}}^{\mu\nu}(\mathcal{T}_\text{H})_{\mu\nu}.
\end{equation}
If $(\textbf{g},\mathcal{T})$ satisfies the NEC or SEC, then so does $(\textbf{g}_\epsilon,\mathcal{T}_\epsilon)$ for each $\epsilon > 0$. From the limits (\ref{limeps}) we see that the energy conditions also hold for $(\textbf{g}_{\text{H}},\mathcal{T}_{\text{H}})$.
\end{proof}
\section{Rigidity theorem} \label{rig}
The proof of Theorem \ref{thm1} involves an Ansatz for the Killing vector and relies on a generalisation of the divergence identity in \cite{DL} (see also the extensions in \cite{CKL, KL24, CD25}). Although we follow the derivation in \cite{DL} below, we also explain how to deduce the identity directly from the Einstein equations for the near-horizon geometry.

\subsection{Divergence identity}  \label{divsec} Given any smooth and strictly positive function $\Gamma$ on the cross-section $M$, we introduce a vector $K$ by
\begin{equation} \label{Kans}
    K^\flat = \Gamma X + \text{d}\Gamma.
\end{equation}
Here $K^\flat$ denotes the 1-form $g$-dual to $K$. Using this relation to eliminate $X = \Gamma^{-1}(K^\flat - \text{d}\Gamma)$, the horizon equations (\ref{nhe}) can be written in terms of $K$ and $\Gamma$ as 
\begin{equation} \label{nhek}
        R_{ab} = \frac{K_aK_b}{2\Gamma^2} - \frac{(\nabla_a\Gamma)(\nabla_b\Gamma)}{2\Gamma^2} - \frac{1}{\Gamma}\nabla_{(a}K_{b)} + \frac{1}{\Gamma}\nabla_a\nabla_b\Gamma + P_{ab},
\end{equation}
where $P_{ab} = T_{ab} - \frac 1n(g^{cd}T_{cd} + 2U)g_{ab}$ represents the matter terms. It will also be useful to define a function $A$ by 
\begin{equation} \label{A}
    A = \Gamma F - \frac{\vert K \vert^2}{\Gamma}.
\end{equation}
We can express the relations (\ref{c1}, \ref{c2}, \ref{F}) in terms of $K,\Gamma$ as 
\begin{subequations} \label{rewritek}
    \begin{align}
        \Gamma \beta_a &= -\nabla^b(\Gamma T_{ab}) + K^bT_{ab} - UK_a + U\nabla_a \Gamma, \label{beta}\\
        \Gamma^2\alpha &= \Gamma K^a\beta_a - \frac 12\nabla_a(\Gamma^2 \beta^a),\label{alpha} \\
        A &= -\frac{\vert K \vert^2}{2\Gamma} + \frac 12 \Delta \Gamma - \frac 12 \nabla_b K^b - \frac{1}{2\Gamma}K^b\nabla_b \Gamma + (1-\tfrac 2n)\Gamma U - \frac 1n\Gamma g^{cd}T_{cd}. \label{A2}
\end{align}
\end{subequations}
Here $\Delta = \nabla^a\nabla_a$ is the Laplacian. The generalisation of the vacuum identity in \cite{DL} reads as follows.

\begin{prop} \label{prop1}
    Suppose the horizon data $(g,K,\Gamma,T,U)$ solves the horizon equations \eqref{nhek}. Then the following identity holds on $M$.
    \begin{align}  \label{magid}
        \frac 14\vert \mathcal{L}_K g \vert^2 + \gamma = \:\:&\nabla^a\left(K^b\nabla_{(a}K_{b)} - AK_a -  K_a\nabla_b K^b -\frac{1}{2\Gamma}K_aK^b\nabla_b\Gamma - \frac{\vert K \vert^2}{2\Gamma}K_a - \frac 12\Gamma^2 \beta_a\right) \nonumber \\
        &+ \nabla_bK^b\left(A + \nabla_a K^a + \frac{1}{\Gamma}K^a\nabla_a\Gamma\right).
    \end{align}
    Here $\alpha,\beta,A$ are given by (\ref{rewritek}) and 
    \begin{equation} \label{Ldef}
        \gamma = T_{ab}K^aK^b - 2\Gamma K^a\beta_a - \vert K \vert^2 U + \Gamma^2 \alpha.
    \end{equation}
\end{prop}
\begin{proof}
    The first part of the proof proceeds as in the vacuum case. We write
    \begin{equation*}
        \frac 14\vert \mathcal{L}_K g\vert^2 = \nabla_{(a}K_{b)}\nabla^aK^b = \nabla^a(K^b\nabla_{(a}K_{b)}) - K^b\nabla^a\nabla_{(a}K_{b)}
    \end{equation*}
    and use the contracted Bianchi identity $\nabla^a(R_{ab}-\frac 12 Rg_{ab}) = 0$ applied to (\ref{nhek}) contracted with $\Gamma K^b$ to compute the last term. The matter content contributes an extra term $-\Gamma K^b\nabla^a(P_{ab}-\frac 12(g^{cd}P_{cd})g_{ab})$ compared to the vacuum calculation. Subsequently, we use the Ricci identity 
    \begin{equation*}
        \Delta\nabla_b\Gamma = \nabla_b\Delta\Gamma + R_{ab}\nabla^a\Gamma
    \end{equation*}
    and (\ref{nhek}) again to rewrite a triple derivative of $\Gamma$. This step introduces an additional matter term $-P_{ab}K^b\nabla^a\Gamma$. Putting everything together,
        \begin{align*} 
        \frac 14\vert \mathcal{L}_K g \vert^2  = \:  &\nonumber\nabla_bK^b\left(-\frac{1}{2\Gamma}\vert K \vert^2 + \frac 12\Delta \Gamma + \frac 12\nabla_b K^b + \frac{1}{2\Gamma}K^b\nabla_b\Gamma \right) \\[3pt]&+\nabla^a\left(K^b\nabla_{(a}K_{b)} - \tfrac 12K_a\Delta \Gamma-\tfrac 12K_a\nabla_b K^b\right) - \Gamma K^a\nabla^b(P_{ab} - \tfrac 12(g^{cd}P_{cd})g_{ab}) - P_{ab}K^a\nabla^b\Gamma.
    \end{align*}
    Denote the matter terms on the second line by $S$. Plugging in the definition of $P_{ab}$ and using (\ref{beta}), 
    \begin{align*}
        S &= -K^a\nabla^b(\Gamma T_{ab}) - \Gamma K^a\nabla_a U + K^a\nabla_a\left(\tfrac 1n\Gamma g^{cd}T_{cd} + \tfrac 2n \Gamma U\right) \nonumber\\
        &= \Gamma K^a\beta_a - K^aK^bT_{ab} +  \vert K \vert^2U  - \nabla_a\left(((1-\tfrac 2n)\Gamma U - \tfrac 1n \Gamma g^{cd}T_{cd})K^a\right) \\&\hspace{1cm}+ \nabla_a K^a \left((1-\tfrac 2n)\Gamma U - \tfrac 1n \Gamma g^{cd}T_{cd}\right).
    \end{align*}
Rearranging and using (\ref{A2}), the terms proportional to $\nabla_b K^b$ become exactly as in (\ref{magid}). The last step to recover the expression for $\gamma$ in (\ref{Ldef}) is to use (\ref{alpha}), which contributes the final divergence term $-\frac 12\nabla_a (\Gamma^2 \beta^a)$ in the identity.
\end{proof}
Alternatively, (\ref{magid}) may be obtained directly from the Einstein equations for the associated near-horizon geometry as follows. In \cite{KL13} the Einstein tensor $\mathcal{G}_{\text{H}}$ of the near-horizon geometry is expressed in terms of the data $(g,X,F)$ that determines $\textbf{g}_\text{H}$ via (\ref{nhga}). Using these formulae and writing $(X,F)$ in terms of $(K,\Gamma,A)$, in the vacuum case ($\mathcal{T}_{\text{H}} = 0$) the identity (\ref{magid}) can be shown to be equivalent to
\begin{equation} \label{idint}
\mathcal{G}_{\textup{H}}(-\Gamma e_++ rK^ie_i, r\Gamma^{-1}\vert K \vert^2e_- + K^ie_i) = rK^iK^j(\mathcal{G}_{\text{H}})_{ij} - r\vert K \vert^2(\mathcal{G}_{\text{H}})_{+-} - \Gamma K^i(\mathcal{G}_{\text{H}})_{i+} = 0,
\end{equation}
where all components are taken in the basis (\ref{onbdef}, \ref{onbdefvec}). The general identity (\ref{magid}) with matter is a linear combination of (\ref{alpha}) (coming from conservation of $\mathcal{T}_\text{H}$) and (\ref{idint}) (with the corresponding components of $\mathcal{T}_\text{H}$ on the right hand side).

\subsection{Horizon rigidity} From now on we assume the associated near-horizon geometry satisfies the null energy condition. At any point $p \in M$, consider the quadratic form $Q$ on $\R \times T_pM$ defined by
\begin{equation}
    Q(\eta, L) = T_{ab}L^aL^b - 2\eta L^a\beta_a - \vert L \vert^2U + \alpha\eta^2.
\end{equation}
When $\eta \neq 0$ we have $r^2Q(\eta,L) = \mathcal{T}_\text{H}(\ell, \ell)$ for a null vector $\ell$ in the near-horizon geometry, expressed in the null-orthonormal frame (\ref{onbdefvec}) as
\begin{equation} \label{ell}
    \ell= \eta e_+ - r L^ie_i - \frac{1}{2\eta}r^2\vert L\vert^2e_-.
\end{equation}
Hence the null energy condition implies $Q$ is positive semi-definite (for $\eta = 0$ this follows by continuity). In particular, the function $\gamma = Q(\Gamma,K)$ in \eqref{Ldef} is pointwise non-negative. We deduce Theorem \ref{thm1} as follows.
\begin{proof}[Proof of Theorem \ref{thm1}]
Observe that until this point $\Gamma$ was an arbitrary smooth positive function. It is proven in \cite{DL} (see also \cite{G84, AMS08, LR12}) that there exists a unique (up to scale) choice of $\Gamma > 0$ such that $K$ is divergence-free, i.e. so that $\Gamma$ solves
\begin{equation} \label{gammadef}
    \Delta \Gamma  +\nabla_a(\Gamma X^a) = 0.
\end{equation}
For this choice of $\Gamma$, the last term in (\ref{magid}) vanishes. Since $\gamma$ is non-negative assuming the null energy condition, integrating\footnote{Here we apply the divergence theorem, which is valid even if $M$ is not orientable. Alternatively, in the non-orientable case we can pass to the orientation cover and argue as in \cite{DL} that $K$ is a Killing vector of $(M,g)$.} (\ref{magid}) over the compact manifold $M$ shows that $\mathcal{L}_K g = 0 = \gamma$. The vector $K$ vanishes if and only if $X = -\text{d}(\log \Gamma)$ is exact, so for rotating solutions we deduce that $K$ is a Killing vector of $(M,g)$.

It remains to prove that $K$ preserves the data $(X,T,U)$. Since $Q$ is positive semi-definite and $Q(\Gamma,K) = 0$, it follows that $(\Gamma,K)$ lies in the kernel of the corresponding $(1+n) \times (1+n)$ symmetric matrix, 
\begin{equation*}
    \begin{bmatrix}
        \alpha & -\beta_a \\ -\beta_a & T_{ab} - Ug_{ab}
    \end{bmatrix} \begin{bmatrix} \Gamma \\ K^a \end{bmatrix} = 0.
\end{equation*}
We infer 
 \begin{equation} \label{simp1}
        \Gamma \alpha = K^a \beta_a, \hspace{.8cm} \Gamma \beta_a + U K_a = K^bT_{ab}.
    \end{equation}
    Using the relations (\ref{beta}, \ref{alpha}) and the fact that $K$ is Killing, we find the horizon data satisfies
    \begin{equation} \label{simp2}
        K^a\nabla^b(\Gamma T_{ab}) = K^a\nabla_a(\Gamma U), \hspace{.8cm} \nabla^b(\Gamma T_{ab}) = U\nabla_a \Gamma.
    \end{equation}
    Contracting the second equation with $K$ and comparing to the first shows $\mathcal{L}_K U = 0$. To prove $K$ preserves $\Gamma$, we go back to (\ref{nhek}) and argue as in \cite{CDKL} (see also \cite{GSW25}). The trace of (\ref{nhek}) reads
        \begin{equation*}
        R = \frac{\vert K \vert^2}{2\Gamma^2} - \frac{\vert \nabla \Gamma \vert^2}{2\Gamma^2} + \frac{1}{\Gamma}\Delta \Gamma - 2U.
    \end{equation*}
    Lie-deriving this identity along $K$ using the facts that $\mathcal{L}_K R = \mathcal{L}_K U = 0$, we find  $L(\mathcal{L}_K\Gamma) = 0$, where $L$ is the linear elliptic operator
    \begin{equation}
        L\psi = -\Delta \psi + \nabla_a((\Gamma^{-1}\nabla^a\Gamma)\psi) + \Gamma^{-2}\vert K \vert^2 \psi.
    \end{equation}
    It is proven in \cite{CDKL} that the kernel of $L$ is trivial assuming compactness of $M$. Therefore $\mathcal{L}_K \Gamma = 0$, which implies $\mathcal{L}_K X = 0$. It remains to show $\mathcal{L}_K T = 0$. The Lie derivative of (\ref{nhek}) reduces to
    \begin{equation*}
        \mathcal{L}_K T = \frac 1n\mathcal{L}_K(g^{ab}T_{ab})g.
    \end{equation*}
    Lie-deriving the second equation in (\ref{simp2}) then shows d$(\Gamma \mathcal{L}_K (g^{ab}T_{ab})) = 0$. It follows that $\mathcal{L}_K (g^{ab}T_{ab})$ equals a constant times $\Gamma^{-1}$, and an integration over $M$ shows this constant must be zero. We thus conclude that $\mathcal{L}_K T = 0$.
\end{proof}

Note that the argument for the existence of the Killing vector $K$ requires only the integrated energy condition $\int_M \mathcal{T}_{\text{H}}(\ell,\ell)\text{ vol}_g \geq 0$. The pointwise null energy condition is used to prove that $K$ preserves not just the induced metric $g$ but also the remaining horizon data $(X,T,U)$. 
   
\section{Symmetry enhancement of the near-horizon geometry}  \label{enhsec}
Assuming the NEC, the Killing vector $K$ constructed in Section \ref{rig} leaves the horizon data invariant and therefore extends to a Killing vector of $\textbf{g}_{\text{H}}$ preserving $\mathcal{T}_{\text{H}}$. Following \cite{KLR, DL}, we show that the near-horizon geometry admits yet another Killing vector. In order for this Killing vector to integrate to a well-defined group action, we construct an extension $(\overline{\textbf{g}}_{\text{H}},\overline{\mathcal{T}}_{\text{H}})$ of the near-horizon geometry to which Theorem \ref{thm2} applies (see \cite[Remark 2.9]{DL}). The function $A$ introduced in (\ref{A}) plays a key role in these arguments. A special case where a further symmetry enhancement occurs is discussed in Section~\ref{ads3}. 

\subsection{Symmetry enhancement} \label{senhsec} Whenever the conclusion in Theorem \ref{thm1} holds, the expression for $A$ in (\ref{A2}) reduces to
\begin{equation} \label{Aclean}
A = -\frac{\vert K \vert^2}{2\Gamma} + \frac 12 \Delta \Gamma + (1-\tfrac 2n)\Gamma U - \frac 1n\Gamma g^{ab}T_{ab}.
\end{equation}
We adopt the approach in \cite{KL24, DL} to prove $A$ must be constant. The results in this section are valid both in the rotating and non-rotating case, the only difference being that the vector $K$ vanishes for non-rotating solutions.

\begin{prop} \label{propA}
    Let $(g,X,T,U)$ be a solution to the horizon equations on a compact and connected manifold $M$ such that the associated near-horizon geometry satisfies the NEC. Then the function $A$ defined by \eqref{A} is constant. If in addition the SEC holds and the solution is rotating, this constant is negative.
\end{prop}
\begin{proof}
We repeat the computation in the proof of Proposition \ref{prop1} without contracting with $K$, but instead using (\ref{simp2}) and the fact that $K$ preserves the horizon data. From (\ref{nhek}) we have 
\begin{align*}
    \Gamma\nabla^aR_{ab} &= -\frac{1}{4\Gamma}\nabla_b \vert K \vert^2 -\frac{1}{2\Gamma}(\Delta \Gamma)\nabla_b\Gamma - \frac{3}{4\Gamma}\nabla_b \vert \nabla \Gamma \vert^2 + \frac{1}{\Gamma^2}\vert \nabla \Gamma \vert^2\nabla_b\Gamma + \Delta \nabla_b\Gamma+ \Gamma \nabla^aP_{ab}, \\[3pt]
    \Gamma\nabla_b R &= \Gamma \nabla_b\left(\frac{\vert K \vert^2}{2\Gamma^2} - \frac{\vert \nabla \Gamma \vert^2}{2\Gamma^2} + \frac{\Delta \Gamma}{\Gamma} + g^{cd}P_{cd}\right).
\end{align*}
Using the definition of $P_{ab}$, the contracted Bianchi identity $\Gamma\nabla^a(R_{ab}-\frac 12R g_{ab}) = 0$ becomes
    \begin{equation} \label{Astep}
        \nabla_a\left(-\frac{\vert K \vert^2}{2\Gamma} - \frac 12\Delta \Gamma -\frac{\vert \nabla \Gamma \vert^2}{2\Gamma}\right) + \Delta \nabla_a \Gamma + \Gamma \nabla^bT_{ab} - \frac 1n\Gamma \nabla_a (g^{cd}T_{cd}) + (1- \tfrac 2n)\Gamma\nabla_a U = 0.
    \end{equation}
    The Ricci identity and equation (\ref{nhek}) imply
    \begin{equation*}
        \Delta \nabla_a \Gamma = \nabla_a \Delta \Gamma + R_{ab}\nabla^b\Gamma = \nabla_a \Delta \Gamma + \nabla_a\left(\frac{\vert \nabla \Gamma \vert^2}{2\Gamma}\right) + T_{ab}\nabla^b\Gamma - \frac 1n(g^{cd}T_{cd} + 2U)\nabla_a \Gamma.
    \end{equation*}
    Inserting this expression into (\ref{Astep}) and using the second equation in (\ref{simp2}) to rewrite the matter term $\nabla^b(\Gamma T_{ab})$, all terms can be written as a total derivative. The result is
    \begin{equation*}
        \text{d}\left(-\frac{\vert K \vert^2}{2\Gamma} + \frac 12 \Delta \Gamma + (1-\tfrac 2n)\Gamma U - \frac 1n\Gamma g^{cd}T_{cd}\right) = 0.
    \end{equation*}
    This is precisely the statement d$A = 0$. It remains to prove that $A < 0$ for rotating near-horizon geometries satisfying the strong energy condition. The argument is based on \cite{KLR,KL13}. Consider the vector $\xi = \partial_v - \partial_r$ in the near-horizon geometry, which is timelike on $\mathcal{H} = \{r = 0\}$. The strong energy condition implies
\begin{equation*}
    0 \leq \mathcal{R}_{\text{H}}(\xi,\xi)= \mathcal{T}_{\text{H}}(\xi,\xi)-\frac 1n (\textbf{g}_{\text{H}})^{\mu\nu}(\mathcal{T}_{\text{H}})_{\mu\nu} \textbf{g}_{\text{H}}(\xi,\xi)\:\overset{\mathcal{H}}{=}\: -2(1-\tfrac 2n)U + \frac 2n g^{cd}T_{cd}.
\end{equation*}
Hence the sum of the final two matter terms in the expression for $A$ is non-positive. Integrating $A$, we find
\begin{equation*}
    A\:\text{vol}(M) = \int_M A \text{ vol}_g = \int_M\left(-\frac{\vert K \vert^2}{2\Gamma} + (1-\tfrac 2n)\Gamma U - \frac 1n\Gamma g^{cd}T_{cd}\right)\text{vol}_g < 0,
\end{equation*}
since $K$ is non-zero and the integrand is non-positive.
\end{proof}
Let us return to the near-horizon geometry and introduce a coordinate $\rho$ by $r = \Gamma \rho$. Expressing $F$ and $X$ in terms of $A,K,\Gamma$ using (\ref{A}), we have
\begin{equation} \label{nhgenh}
    \textbf{g}_{\text{H}} = \Gamma(2\text{d}v\text{d}\rho + A\rho^2\text{d}v^2) + 2 K^\flat \rho\text{d}v + \vert K \vert^2 \rho^2\text{d}v^2 + g.
\end{equation}
If $A$ is constant, the two-dimensional metric in the round brackets is maximally symmetric with scalar curvature $2A$. Moreover, as shown in \cite{DL}, in addition to $K, \partial_v$ and $v\partial_v - \rho\partial_\rho$ the near-horizon metric admits a Killing vector
\begin{equation} \label{mkv}
    m= \frac12Av^2\partial_v + (1-A\rho v)\partial_\rho - vK.
\end{equation}
The integral curves of $m$ are not complete if $A \neq 0$, as $\vert v \vert \to \infty$ in finite parameter time due to the term $v^2\partial_v$. In order for $m$ to integrate to an isometric $\R$-action, we need to extend the $\R^2$ factor of the near-horizon geometry to a surface $\Sigma$ which is either the global AdS$_2$ spacetime or global dS$_2$, depending on whether $A$ is negative or positive respectively. Let us write $A = \varepsilon\kappa^{-2}$ with $\varepsilon = +1$ for the dS$_2$ case and $\varepsilon = -1$ for AdS$_2$. We can view $\Sigma$ as a hyperboloid
\begin{equation*}
    X_2^2 - X_1^2-X_0^2 = -\kappa^2,
\end{equation*}
embedded in $\R^{3}$ with metric $\varepsilon(\text{d}X_0^2  +\text{d}X_1^2 - \text{d}X_2^2)$. The relation to $(\rho,v)$ coordinates is 
\begin{equation} \label{hyp1}
    X_0 +X_2 = \rho, \hspace{.8cm} X_1 = \kappa^{-1}\rho v -\varepsilon\kappa, \hspace{.8cm} X_2 - X_0 = \kappa^{-2}\rho v^2-2\varepsilon v.
\end{equation}
Note that these coordinates cover the whole hyperboloid with the exception of the line where $X_1 = \varepsilon\kappa$ and $X_0 + X_2 = 0$. We now introduce global coordinates $(\tau,\sigma)$ on $\Sigma$ by
\begin{equation} \label{hyp2}
    X_0 = \sqrt{\kappa^2 + \sigma^2}\cos\tfrac{\tau}{\kappa}, \hspace{.8cm} X_1 = \sqrt{\kappa^2 + \sigma^2}\sin \tfrac{\tau}{\kappa}, \hspace{.8cm} X_2 = \sigma.
\end{equation}
In the AdS$_2$ case we may pass to the universal cover of $\Sigma$ to avoid closed timelike curves, whereas in the dS$_2$ case $\tau$ is periodic. Comparing (\ref{hyp1}) to (\ref{hyp2}),
\begin{equation}
    (v,\rho) = \Omega(\tau,\sigma) = \left(\frac{\kappa\sqrt{\kappa^2+\sigma^2}\sin\frac{\tau}{\kappa} +\varepsilon\kappa^2}{\sqrt{\kappa^2 + \sigma^2}\cos\frac{\tau}{\kappa} + \sigma},\sigma + \sqrt{\kappa^2 + \sigma^2}\cos \tfrac{\tau}{\kappa}\right).
\end{equation}
This transformation satisfies $\text{d}v\wedge\text{d}\rho = \text{d}\tau \wedge\text{d}\sigma$, from which we deduce the existence of a function $\zeta(v,\rho)$ on $\mathbb{R}^2$ such that $\sigma\text{d}\tau = \rho\text{d}v + \text{d}\zeta$. In the rotating case we supplement the coordinate transformation with a flow $\Psi^K_\zeta$ for time $\zeta(v,\rho)$ along the integral curves of $K$. This satisfies\footnote{In local coordinates $(y^i,\chi)$ on $M$ such that $K = \partial_\chi$, we can write (\ref{nhgenh}) as 
\begin{equation*}
        \textbf{g}_{\text{H}} = \Gamma(2\text{d}v\text{d}\rho + A\rho^2\text{d}v^2) + g_{\chi\chi}(\text{d}\chi +\rho\text{d}v)^2 + 2g_{i\chi}\text{d}y^i(\text{d}\chi + \rho\text{d}v) + g_{ij}\text{d}y^i\text{d}y^j.
\end{equation*}
Here $\Gamma$ and the metric components depend on the $y^i$ only. The transformation $\Psi^K_\zeta$ corresponds to a shift $\chi \mapsto \chi + \zeta$, which ensures that $\text{d}\chi +\rho\text{d}v \mapsto \text{d}\chi + \sigma\text{d}\tau$.}
\begin{equation} \label{comp}
    (\Psi^K_\zeta)^* K^\flat  = K^\flat + \vert K \vert^2\text{d}\zeta, \hspace{.8cm}(\Psi^K_\zeta)^* g = g + 2K^\flat \text{d}\zeta + \vert K \vert^2 \text{d}\zeta^2.
\end{equation}
Setting $\Omega(\tau,\sigma) = (\tau,\sigma)$ for $A = 0$ and $f = 1 + \frac{\sigma^2}{\kappa^2}$, altogether we have
\begin{align}
    (\Omega \circ \Psi_\zeta^K)^*\textbf{g}_{\text{H}} = \begin{cases}
        \Gamma\varepsilon\left(-f^{-1}\text{d}\sigma^2 +f\text{d}\tau^2\right)  + 2K^\flat\sigma\text{d}\tau + \vert K \vert^2 \sigma^2\text{d}\tau^2 + g &\text{if   } A \neq 0, \\[5pt]
        2\Gamma\text{d}\tau\text{d}\sigma + 2K^\flat \sigma\text{d}\tau + \vert K \vert^2 \sigma^2\text{d}\tau^2 + g &\text{if   } A =0.
    \end{cases}
\end{align}
In the new coordinates we can extend the near-horizon geometry to all values of $\tau, \sigma \in \R$ to obtain the extension $(\Sigma \times M,\overline{\textbf{g}}_{\text{H}})$ to which Theorem \ref{thm2} applies (with $\Sigma = \mathbb{R}^2$ if $A = 0$).  The energy-momentum tensor $\mathcal{T}_{\text{H}}$ can similarly be extended to a tensor $\overline{\mathcal{T}}_{\text{H}}$ on $\Sigma \times M$, because using (\ref{simp1}) we can write it as
\begin{equation} \label{Tenh}
    \mathcal{T}_{\text{H}} = \Gamma U(2\text{d}v\text{d}\rho + A\rho^2\text{d}v^2) + 2(\iota_K T) \rho\text{d}v + T_{ab}K^aK^b\rho^2\text{d}v^2 + T.
\end{equation}
\begin{proof}[Proof of Theorem \ref{thm2}] 
Let $\Phi$ be an element of the identity component of the isometry group $G$ of AdS$_2$, 2D Minkowski space or dS$_2$ depending on whether $A$ is negative, zero or positive respectively. In the extended near-horizon geometry $\Phi$ has a well-defined action on $\Sigma$. Since $\Phi$ preserves the volume form d$\tau \wedge \text{d}\sigma$, the 1-form $\Phi^*(\sigma\text{d}\tau) - \sigma\text{d}\tau$ is closed. In fact, there exists a globally defined function $H_\Phi$ on $\Sigma$ such that
\begin{equation} \label{Hpsi}
    \text{d}H_\Phi = \Phi^*(\sigma\text{d}\tau) - \sigma\text{d}\tau.
\end{equation}
Indeed, let $\Psi_t$ for $t \in [0,1]$ be a path in $G$ with $\Psi_0 = \text{Id}$ and $\Psi_1 = \Phi$, and let $\xi_t$ be the Killing vector field generating $\Psi_t$. It is straightforward to verify that there exists a global\footnote{This is immediate for $A = 0$ since $\Sigma$ is simply connected. When $A \neq 0$, a basis of Killing vectors on $\Sigma$ is
\begin{equation*}
    k_1 = \kappa\partial_\tau, \hspace{.5cm} k_2 = \sqrt{\kappa^2 + \sigma^2}\cos\tfrac{\tau}{\kappa}\partial_\sigma - \frac{\kappa\sigma\sin\tfrac{\tau}{\kappa}}{\sqrt{\kappa^2 + \sigma^2}}\partial_\tau, \hspace{.5cm} k_3 = \sqrt{\kappa^2 + \sigma^2}\sin\tfrac{\tau}{\kappa}\partial_\sigma + \frac{\kappa\sigma\cos\tfrac{\tau}{\kappa}}{\sqrt{\kappa^2 + \sigma^2}}\partial_\tau.
\end{equation*} These satisfy
\begin{equation*}
    \mathcal{L}_{k_1} (\sigma\text{d}\tau) = 0, \hspace{.7cm} \mathcal{L}_{k_2} (\sigma\text{d}\tau) = \text{d}\left(\frac{\kappa^3\sin\tfrac{\tau}{\kappa}}{\sqrt{\kappa^2 + \sigma^2}}\right), \hspace{.7cm}  \mathcal{L}_{k_3} (\sigma\text{d}\tau) = \text{d}\left(-\frac{\kappa^3\cos\tfrac{\tau}{\kappa}}{\sqrt{\kappa^2 + \sigma^2}}\right).
\end{equation*}} function $h_{\xi_t}$ such that $\mathcal{L}_{\xi_t} (\sigma\text{d}\tau) = \text{d}h_{\xi_t}$. We have
\begin{equation*}
    \Phi^*(\sigma\text{d}\tau) - \sigma\text{d}\tau = \int_0^1\Psi_t^*(\mathcal{L}_{\xi_t} (\sigma\text{d}\tau))\:\text{d}t = \int_0^1\Psi_t^*(\text{d}h_{\xi_t})\:\text{d}t = \text{d}\left[\int_0^1\Psi_t^*h_{\xi_t} \:\text{d}t\right].
\end{equation*}
Therefore we may take
\begin{equation*}
    H_\Phi(\tau,\sigma) = \int_0^1 h_{\xi_t}(\Psi_t(\tau,\sigma))\:\text{d}t.
\end{equation*}
This definition may depend on the choice of path, but only by an additive constant as (\ref{Hpsi}) is satisfied. We may fix this ambiguity by requiring $H_\Phi(0,0) = 0$. We now extend $\Phi$ to an isometry  $\overline{\Phi}$ of the extended near-horizon geometry by setting, for $x \in M$,
\begin{equation*}
    \overline{\Phi}((\tau,\sigma), x) = (\Phi(\tau,\sigma), \Psi^K_{-H_\Phi}(x))
\end{equation*} 
Here $\Psi^K_{-H_\Phi}$ denotes the flow along $K$ for time $-H_\Phi(\tau,\sigma)$. The same computation as in (\ref{comp}) shows this is an isometry of $\overline{\textbf{g}}_{\text{H}}$. We hence obtain a faithful isometric action of the identity component of $G$. This can be extended to the orientation-preserving subgroup of $G$ by noting that the discrete isometry $(\sigma,\tau) \mapsto (-\sigma,-\tau)$ preserves $\overline{\textbf{g}}_{\text{H}}$. It follows from (\ref{Tenh}) that $\overline{\mathcal{T}}_{\text{H}}$ admits the same symmetry enhancement as $\overline{\textbf{g}}_{\text{H}}$. If the strong energy condition holds and the horizon data is rotating we have $A < 0$ by Proposition \ref{propA}, so we are in the AdS$_2$ case.
\end{proof}

\subsection{\texorpdfstring{$\text{AdS}_3$}{AdS3} near-horizon geometries}  \label{ads3} It is possible for an extremal horizon to be both static and rotating according to the definitions in Section \ref{ehors}. An explicit example of a black hole containing such a horizon is the supersymmetric black ring \cite{EEMR04} in five-dimensional minimal supergravity. The near-horizon geometry in this case is a direct product of a round $S^2$ with a 3D space locally isometric to AdS$_3$. In this section we generalise the arguments in \cite{KLR} to show that horizons that are both rotating and static admit a further symmetry enhancement and can locally be written as a warped product with AdS$_3$.

\begin{prop} \label{propads3}
Consider a near-horizon geometry as in Theorem \ref{thm2} whose associated horizon data is both static and rotating. Then $A < 0$ and the metric $\textup{\textbf{g}}_{\textup{H}}$ can locally be written as a warped product of a base manifold $N$ with \textup{AdS}$_3$. In particular, the Lie algebra of the isometry group contains a subalgebra $\mathfrak{so}(2,2)$ that preserves $\mathcal{T}_{\textup{H}}$.
\end{prop}
\begin{proof}
    For rotating and static horizons the 1-form $X$ is closed but not exact, and d$F = XF$ with $F$ as in (\ref{F}). In this case $K$ is non-zero and $\Gamma^{-1}K^\flat$ is closed. Since $K$ is Killing and $\mathcal{L}_K \Gamma = 0$, it follows that $K$ is parallel with respect to the rescaled metric $\Gamma^{-1}g$. This implies that, at least locally, $M$ splits isometrically as a product $\R \times N$. Moreover, as $\Gamma^{-1}\vert K \vert^2$ is constant, equation (\ref{A}) shows $F$ equals a constant $c$ times $\Gamma^{-1}$. The condition d$F = XF$ ensures that $c = 0$, as otherwise $X$ would be exact. From (\ref{A}) we now find $\vert K \vert^2 = -A\Gamma$ (in particular, we must have $A < 0$). Hence, choosing a coordinate $\chi$ on $\R$ such that $K = \partial_\chi$, we locally have
    \begin{equation} \label{wp}
        g = -A\Gamma \text{d}\chi^2 + g_N,
    \end{equation}
    with $\Gamma$ a positive function on $(N,g_N)$. The near-horizon metric (\ref{nhgenh}) becomes
    \begin{equation} \label{gads3}
        \textbf{g}_{\text{H}} = \Gamma(2\text{d}v\text{d}\rho + A\rho^2\text{d}v^2) -A\Gamma(\text{d}\chi + \rho\text{d}v)^2 + g_N = \Gamma(2\text{d}\rho\text{d}v - 2A\rho\text{d}\chi\text{d}v - A\text{d}\chi^2) + g_N.
    \end{equation}
    The 3D metric in the final brackets is locally isometric to AdS$_3$ (see (\ref{ads3c})). Hence, in this case the near-horizon geometry is locally a warped product of $N$ with AdS$_3$ and admits a six-dimensional space of Killing vectors forming the Lie algebra $\mathfrak{so}(2,2)$. To see that $\mathcal{T}_{\text{H}}$ admits the same symmetry enhancement, we can argue that the Einstein tensor $\mathcal{G}_{\text{H}}$ can also be written in the warped product form (\ref{gads3}) and then use the Einstein equations for the near-horizon geometry. Equivalently, this may be deduced from the horizon equations as follows. Consider the identity
    \begin{equation*}
        R_{ab}K^a = [\nabla_a,\nabla_b]K^a = \nabla_a\nabla_bK^a = K_b\left(\frac{\vert \nabla \Gamma \vert^2}{2\Gamma^2} - \frac{\Delta \Gamma}{2\Gamma}\right).
    \end{equation*}
    From the horizon equations (\ref{nhek}) we obtain 
    \begin{equation*}
        R_{ab}K^a = K_b\left(\frac{\vert K \vert^2}{2\Gamma^2} + \frac{\vert \nabla \Gamma \vert^2}{2\Gamma^2} - \frac 2n U - 
        \frac 1n g^{cd}T_{cd}\right) + T_{ab}K^a.
    \end{equation*}
    Combining these equations with (\ref{Aclean}) and the fact that $\vert K \vert^2 = - A\Gamma$, we find $T_{ab}K^b = UK_a$. In particular, (\ref{simp1}) shows that $\alpha$ and $\beta$ both vanish. Writing $T_N$ for the restriction of $\mathcal{T}_{\text{H}}$ to $N$, it follows that $\mathcal{T}_{\text{H}}$ is of the warped product form
    \begin{equation}
        \mathcal{T}_{\text{H}} = U\Gamma(2\text{d}\rho\text{d}v - 2A\rho\text{d}\chi\text{d}v - A\text{d}\chi^2) + T_N,
    \end{equation}
    which is invariant under $\mathfrak{so}(2,2)$.
\end{proof}
As far as we are aware, for all known regular static and rotating near-horizon geometries the function $\Gamma$ is a constant, so that the solution is in fact a direct product of AdS$_3$ with $N$. Examples with non-constant $\Gamma$ have been constructed, but they suffer from conical singularities (see e.g. the examples in five-dimensional $U(1)^3$-gauged supergravity \cite{KL07} or in five-dimensional Einstein-Maxwell theory \cite{KL09s}). It would be interesting to either construct or exclude regular non-trivial warped product AdS$_3$-geometries.

To convert Proposition \ref{propads3} into a statement about isometries as in Theorem \ref{thm2}, one would have to take the universal cover of the near-horizon geometry, transform to global coordinates on the 3D factor in (\ref{gads3}) and then extend the near-horizon geometry to a spacetime of the form AdS$_3 \times \tilde{N}$ (with $\tilde{N}$ simply connected). Setting $A = -\kappa^{-2}$ and viewing AdS$_3$ as a hyperboloid in $\R^4$,
\begin{equation*}
    X_3^2 + X_2^2 - X_1^2 - X_0^2 = -\kappa^2,
\end{equation*}
the relation to global coordinates can be obtained explicitly from
\begin{equation} \label{ads3c}
\begin{aligned}
    X_0 &= 2\kappa^2(v+1)e^{-\frac{\chi}{2\kappa^2}} -\tfrac 14\rho\kappa^{-2}e^{\frac{\chi}{2\kappa^2}}, \\
    X_1 &= \kappa e^{-\frac{\chi}{2\kappa^2}} + \tfrac 12(2\kappa + \kappa^{-1}\rho(v+1))e^{\frac{\chi}{2\kappa^2}}, \\
    X_2 &= \kappa e^{-\frac{\chi}{2\kappa^2}} -  \tfrac 12(2\kappa + \kappa^{-1}\rho(v+1))e^{\frac{\chi}{2\kappa^2}}, \\
    X_3 &=  -2\kappa^2(v+1)e^{-\frac{\chi}{2\kappa^2}} -\tfrac 14\rho\kappa^{-2}e^{\frac{\chi}{2\kappa^2}}. 
\end{aligned}
\end{equation}
The extended near-horizon geometry has isometry group containing $O(2,2)$. We show in Sections \ref{fsec} and \ref{gsec} that matter fields are invariant under the orientation-preserving subgroup  $SO(2,2)$.

\section{Doubly degenerate horizons}  \label{ddhbigsec}
In the presence of a positive cosmological constant, the SEC may be violated and the constant $A$ can become zero or positive. In this section we show that $A$ vanishes for ``triple horizon" configurations like the ultracold Reissner-Nordström-de Sitter horizon. The implications for the Aretakis instability on a background containing such a horizon are discussed in Section \ref{secare}. We assume that the NEC is satisfied, so that Theorems \ref{thm1} and \ref{thm2} apply.

\subsection{Horizons with \texorpdfstring{$A = 0$}{A = 0}}\label{ddhsec} When the Cauchy horizon, event horizon and cosmological horizon coincide in the Reissner-Nordström-dS spacetime, the resulting horizon $\mathcal{H}$ is doubly degenerate in the sense that the norm $\textbf{g}(k,k)$ of the generator $k$ has not just a double, but a triple zero on $\mathcal{H}$. Such triple horizon configurations are also possible for rotating black holes like Kerr-dS. However, in the non-extremal Kerr-dS spacetime the three horizons are not generated by the same Killing vector and therefore the function $\textbf{g}(k,k)$ does not have a triple zero when the horizons merge. Below we construct a different function accounting for rotation that vanishes on each Killing horizon, from which it becomes clear that $A = 0$ for doubly degenerate configurations. The construction uses a vector field $V$ that naturally appears in the context of the weak rigidity theorem \cite{C69, H96,CC08}.

Consider an extremal horizon $\mathcal{H}$ with compact cross-section $M$ in a spacetime with isometry group containing $\R \times U(1)^N$,  with generators $\xi$ and  $m_I$ for  $1 \leq I \leq N$. We assume that 
\begin{enumerate}
 \renewcommand{\labelenumi}{(A\arabic{enumi})}
  \renewcommand{\theenumi}{A\arabic{enumi}}
    \item Any Killing horizon in the spacetime is generated by a Killing vector of the form $k = \xi - \Omega^I_{\mathcal{H}}m_I$ for some constants $\Omega^I_{\mathcal{H}}$. \label{a1}
    \item \label{a2} On $M$, the $m_I$ are tangent to $M$ and the vector $K$ constructed in Section \ref{rig} is a linear combination of the $m_I$.
\end{enumerate}
In particular, the $m_I$ are spacelike or zero at least in a neighbourhood of the horizon. Where it is defined, consider the vector field
\begin{equation}  \label{Vdef}
    V = \xi - \Omega^I m_I, \hspace{.8cm} \text{where } \hspace{.2cm}\Omega^I = h^{IJ}\textbf{g}(\xi,m_J), \hspace{.4cm} h_{IJ} = \textbf{g}(m_I,m_J).
\end{equation}
Here $h^{IJ}$ is the inverse of $h_{IJ}$. Note that the $\Omega^I$ are functions on the spacetime. As $V$ is (up to scale) the unique vector in the span of $\xi,m_I$ that is orthogonal to all $m_I$, on any Killing horizon $\mathcal{H}$ it must be equal to the generator $k$ of $\mathcal{H}$ by our assumption (\ref{a1}). In particular, $\textbf{g}(V,V)$ vanishes and the $\Omega^I = \Omega^I_{\mathcal{H}}$ are constant on $\mathcal{H}$.

Let us introduce the coordinate $\rho = \Gamma^{-1} r$ near $\mathcal{H}$ by rescaling the Gaussian null coordinate $r$ as we did for the near-horizon geometry in (\ref{nhgenh}). Here the function $\Gamma$, initially defined on a cross-section, is extended to a Gaussian null coordinate chart in any way such that it is strictly positive and $\mathcal{L}_k\Gamma = 0$. When two or three Killing horizons coincide, the function $\textbf{g}(V,V)$ will have a zero of order 2 or 3 respectively in $\rho$ on $\mathcal{H}$. The quadratic term in $\rho$ can be calculated using the near-horizon geometry, 
\begin{equation}
    \textbf{g}(V,V) = \textbf{g}_\text{H}(V_\text{H},V_\text{H}) + O(\rho^3).
\end{equation}
Here $V_{\text{H}}$ denotes the vector (\ref{Vdef}) in the near-horizon geometry, which inherits the Killing vectors $\xi, m_I$.
To compute the quadratic term, observe that on the horizon we have $\mathcal{L}_{m_I}\Gamma = 0$. Indeed, as each $m_I$ is a Killing vector commuting with $k$ the function $\mathcal{L}_{m_I}\Gamma$ satisfies the PDE (\ref{gammadef}), so by uniqueness it must be proportional to $\Gamma$. Since $\mathcal{L}_{m_I}\Gamma$ integrates to zero on $M$, the proportionality constant must be zero. Hence, using assumption (\ref{a2}),
\begin{equation} \label{Vnhg}
    V_{\text{H}} = k - \rho K, \hspace{1cm} \textbf{g}_{\text{H}}(V_{\text{H}}, V_{\text{H}}) = A\Gamma \rho^2.
\end{equation}
It follows that $A = 0$ for triple horizons. This motivates the following definition.
\begin{definition} \label{def}
    An extremal horizon $\mathcal{H}$ is \textit{doubly degenerate} if the constant $A$ defined by (\ref{A}) vanishes.
\end{definition}
$A$ behaves like an extremal counterpart of the surface gravity, being a constant that vanishes if the horizon degenerates. Just like for the surface gravity, there is a scaling freedom in the definition of $A$ that can be traced back to the scaling freedom in $\Gamma$. Fixing the normalisation of $A$ requires information extrinsic to the horizon, such as a preferred radial coordinate $\rho$ (e.g. coming from comparison to a Boyer-Lindquist-like radial coordinate, see Appendix \ref{Aa}).

As a consequence of Proposition \ref{propA}, we obtain
\begin{cor}
    In a spacetime satisfying the NEC and the SEC, rotating doubly degenerate horizons cannot exist.
\end{cor}
Note that the assumption that the horizon is rotating cannot be omitted, as for example the flat metric on $\R^{1,1} \times S^1$ contains Killing horizons that are doubly degenerate according to the definition above. Examples of spherically symmetric spacetimes satisfying the dominant energy condition and containing a doubly degenerate horizon were constructed in \cite{KU22}.

If $K$ is of the form $K = \omega^Im_I$ for some constants $\omega^I$, comparing (\ref{Vdef}) to (\ref{Vnhg}) shows that $\partial_\rho (\Omega^I)\vert_{\rho = 0} = \omega^I$. Hence the condition that the horizon is rotating as defined in Section \ref{ehors} corresponds to the requirement that the extensions $\Omega^I$ of the angular velocities $\Omega^I_{\mathcal{H}}$ are not all constant ``to first order" away from $\mathcal{H}$. The usual definition of rotation, which requires knowledge of the asymptotic region to single out the stationary Killing vector, is that the $\Omega^I_{\mathcal{H}}$ are not all zero. These two notions coincide in most cases, since if $\Omega^I$ is non-zero on $\mathcal{H}$ it cannot be constant everywhere as it must vanish asymptotically. However, it is possible for a horizon to be rotating only according to the intrinsic definition, which occurs for example for the supersymmetric black ring in \cite{EEMR04}.
\subsection{Aretakis instability} \label{secare} The multiplicity of the horizon affects the differential order at which the Aretakis instability kicks in, as we now explain following the analysis in \cite{LR12}. Let $\Phi$ be a massless real scalar field satisfying the wave equation on $(\mathcal{M},\textbf{g})$,
\begin{equation} \label{wave}
    \square_{\textbf{g}}\Phi = 0.
\end{equation}
Importantly, in this subsection we do \textit{not} impose that $\Phi$ is invariant under the generator $k$ of the extremal horizon $\mathcal{H}$ in $(\mathcal{M},\textbf{g})$. Initial data is prescribed on a spacelike hypersurface $\Sigma$ whose intersection with $\mathcal{H}$ is a compact cross-section $M$ of the horizon. In the extremal Reissner-Nordstr\"om spacetime, Aretakis showed that $\Phi$ decays along $\mathcal{H}$, assuming an appropriate notion of energy of the initial data is finite. However, the transverse derivative $\partial_r \Phi$ generically does not decay and higher derivatives grow polynomially in the affine parameter $v$ \cite{A11,A112}. Similar results hold for an axisymmetric scalar field on the extremal Kerr spacetime \cite{A113, A12}, and even worse instabilities arise for non-axisymmetric fields \cite{G23}.

We will consider a general extremal horizon, concentrating on the doubly degenerate case as in Definition \ref{def}. Although we make use of properties of the spacetime not determined by the horizon data, we emphasise that all arguments rely only on the geometry in a neighbourhood of $\mathcal{H}$. Starting from Gaussian null coordinates as in Section~\ref{nhgs}, we again introduce the coordinate $\rho = \Gamma^{-1}r$. Unlike in \cite{LR12}, the function $\Gamma = \Gamma(r,x)$ is allowed to depend on both $r$ as well as the coordinates $x^i$, provided it is nowhere-vanishing and agrees at $r = 0$ with the function $\Gamma$ on $M$ constructed in Section \ref{divsec}. The $r$-dependence of $\Gamma$ is partially fixed in Lemma~\ref{lemel}. We also extend the vector $K$ and function $A$ on $M$ constructed in Section \ref{divsec} to a Gaussian null coordinate chart using the components in (\ref{ggnc}), 
\begin{equation} \label{KAext}
      K_i(\rho,x) = \Gamma X_i(\rho,x) + \partial_i\Gamma, \hspace{1cm}   A(\rho,x) = \Gamma F(\rho,x) - \Gamma^{-1}\vert K \vert^2_{g(\rho,x)}.
\end{equation}
Here $\Gamma = \Gamma(\rho,x)$ and all partial derivatives are taken in the chart $(v,\rho,x^i)$. Latin indices are raised and lowered with the induced metric $g_{ij}(\rho,x)$ on the submanifold $M(v,\rho)$ of constant $(v,\rho)$. Observe that on the horizon $K = K^i\partial_i$ is a Killing vector of $g$ and $A$ is constant. 

We can use the freedom in $\Gamma$ (or, equivalently, the radial coordinate $\rho$) to set $\partial_\rho(\nabla_i K^i)$ to zero on the horizon, where $\nabla$ is the induced covariant derivative on $M(v,\rho)$. This is analogous to the way we imposed $\nabla_i K^i = 0$ by solving the PDE (\ref{gammadef}).
\begin{lemma} \label{lemel}
    There exists a choice for the function $\partial_r\Gamma\vert_{r = 0}$ on $M$ such that $\partial_\rho(\nabla_iK^i)$ vanishes on the horizon. Moreover, this function is unique up to an additive constant.
\end{lemma}
\begin{proof}
    On $\mathcal{H}$ we have
\begin{align}
    \partial_\rho(\nabla_i K^i) &= \nabla_i\left((\partial_\rho + \lambda)K^i\right) = \nabla_i\left(\lambda K^i + X^i\partial_\rho\Gamma + \Gamma\partial_\rho X^i + (\partial_\rho g^{ij})\partial_j\Gamma + g^{ij}\partial_j\partial_\rho\Gamma\right) \nonumber \\
    &= \Delta(\Gamma \partial_r\Gamma) + \nabla_i(\Gamma X^i\partial_r\Gamma) + \nabla_i\left(\lambda K^i + \Gamma^2\partial_rX^i + \Gamma(\partial_rg^{ij})\partial_j\Gamma\right). \label{epde}
\end{align}
Here $\lambda = (2\:\text{det }g)^{-1}\partial_\rho(\text{det }g)$ is the expansion along the null vector field $\partial_\rho$ and in the last step we used the fact that $\partial_\rho = \Gamma\partial_r$ on $\mathcal{H}$. Let us denote the final divergence term by $-f$, and observe that $f$ is fixed on $\mathcal{H}$ by prescribing $\Gamma$ on $\mathcal{H}$. The vanishing of (\ref{epde}) is an elliptic PDE for $\psi = (\Gamma\partial_r\Gamma)\vert_{r = 0}$ on $M$, of the form
    \begin{equation} \label{pder}
       L\psi = f, \hspace{.5cm}\text{ where }\hspace{.5cm} L \psi = \Delta \psi + \nabla_i(\psi X^i).
    \end{equation}
    Observe that $L$ is exactly the operator considered in (\ref{gammadef}) in the proof of Theorem \ref{thm1}. In particular, its kernel consists of constant multiples of $\Gamma\vert_{r = 0}$. The formal adjoint of $L$ is $L^*\psi = \Delta \psi - X^i\nabla_i\psi$. By the strong maximum principle (Theorem 2.9 in \cite{K93}) and compactness of $M$, the kernel of  $L^*$ consist of constant functions. Since $f$ is a total divergence, it is orthogonal (in the $L^2$ inner product) to constant functions. It follows using the Fredholm Alternative \cite[Theorem 2.4]{K93} that there exists a (smooth) solution $\psi$ to $L \psi = f$. We may use this solution to fix $\Gamma(r,x)$ to linear order (and hence $\rho$ to quadratic order) in $r$ so that $\partial_\rho(\nabla_i K^i) = 0$ on $\mathcal{H}$. The function $\psi$ is unique up to an element of the kernel of $L$, i.e. up to an additive multiple of $\Gamma\vert_{r = 0}$.
\end{proof}
From now on we fix $\Gamma$ such that it agrees to linear order in $r$ with a function as in Lemma \ref{lemel}. For spherically symmetric spacetimes with $X \equiv 0$ we may take $\Gamma = 1$ and $\rho = r$. Note that the transverse derivative $\partial_\rho$ is always invariant under translations in $v$.

We further define
\begin{equation} \label{I}
    I(v,\rho) = \int_{M(v,\rho)}(2\partial_\rho \Phi + \lambda\Phi)\text{ vol}_g,
\end{equation}
where as before $\lambda$ denotes the expansion along $\partial_\rho$. It is shown in \cite{LR12} that $I = I_0$ is independent of $v$ on $\mathcal{H}$, and, if $AI_0 \neq0$ and $\Phi$ decays as $v \to \infty$, the derivative $\partial_\rho I$ blows up along $\mathcal{H}$. Note that $I_0$ is non-zero for generic initial data. In particular, if $\Phi$ decays as $v \to \infty$, then $\partial_\rho \Phi$ generically does not and a quantity involving $\partial_\rho^2\Phi$ blows up. For doubly degenerate horizons with $A = 0$, we instead find that $\partial_\rho I$ is also conserved and only the second derivative $\partial_\rho^2 I$ generically grows along the horizon, provided the function 
\begin{equation}
    B = \partial_\rho A \vert_{\rho = 0}
\end{equation}
is constant and non-zero. Observe that $B$, unlike $A$, is not determined by the horizon data and hence depends on the spacetime in which the horizon is embedded. If $K$ preserves the first $r$-derivative of the data $(g,X,\Gamma,F)$ on the (doubly degenerate) horizon, it may be verified that $B$ satisfies the following properties:
\begin{enumerate}
    \item $B$ is independent of the extension of $\Gamma$ off the horizon. Explicitly, using (\ref{KAext}) we find
    \begin{equation*}
        B = \Gamma^2 \partial_rF\vert_{r=0} + K^iK^j\partial_rg_{ij}\vert_{r = 0} - 2\Gamma g^{ij}K_i\partial_r X_j\vert_{r =0} - 2K^i\partial_i(\partial_r\Gamma)\vert_{r = 0},
        \end{equation*}
        and the final term vanishes if $K$ preserves $\partial_r\Gamma$.
    \item $B$ is invariant under a change of Gaussian null coordinates (invariant under $K$) corresponding to a different choice of cross-section, see \cite[Equation 16]{LL15}.
    \item $B$ is a non-zero constant for the doubly degenerate Kerr-Newman-de Sitter horizon (see Appendix \ref{Aa} for an explicit expression). We also verified the constancy of $B$ for the five-dimensional doubly degenerate Myers-Perry-de Sitter horizon.
\end{enumerate}
As explained below, $B$ can be thought of as an analogue of $A$ for doubly degenerate horizons. In particular, $B = 0$ in the triply degenerate case. Since the Einstein equations imply that $A$ is constant, it would be interesting to determine whether the same is true for $B$. \\

\noindent \textbf{Open problem}: find an invariant (coordinate-independent) expression for $B$ on a doubly degenerate horizon and determine whether its constancy can be deduced from the Einstein equations.\\

Addressing this problem would likely require proving that first order transverse derivatives of the near-horizon data are invariant under $K$ on the horizon. Assuming that $B$ is constant, the Aretakis instability shifts by one order in the following sense.

\begin{prop} \label{areprop}
    Consider a solution $\Phi$ to the wave equation \eqref{wave} in a spacetime containing a doubly degenerate horizon $\mathcal{H}$.
    \begin{enumerate}[(i)]
        \item \textup{(Non-decay)} Both $I(v,0) = I_0$ and $\partial_\rho I(v,0) = I_1$ are conserved along $\mathcal{H}$, with $I$ as in \eqref{I}.
        \item \textup{(Blow-up)} If $B$ is constant and $BI_0 \neq 0$, then either $\Phi$ does not decay along $\mathcal{H}$ or $\partial_\rho^2 I(v,0)$ blows up linearly in $v$ as $v \to \infty$.
    \end{enumerate}
\end{prop}
\begin{proof}
To compute the wave operator (\ref{wave}) in the coordinates $(v,\rho,x^i)$, we require the inverse metric
\begin{equation} \label{invmet}
    \textbf{g}^{-1} = 2C\partial_v\partial_\rho - \Gamma\rho^2C^{2}A\partial_\rho\partial_\rho -2\rho CK^i\partial_\rho\partial_{x^i} + g^{ij}\partial_{x^i}\partial_{x^j}.
\end{equation}
Here $g^{ij}$ denotes the inverse of the metric $g_{ij}$ on $M(v,\rho)$, the functions $K^i$ and $A$ are given by (\ref{KAext}) and $C^{-1} = \Gamma + \rho\partial_\rho\Gamma$.
All components of (\ref{invmet}) depend on $\rho$ and $x^i$. The wave operator reads
\begin{align}\label{wavecoord}
    0 &= C^{-1}\square_{\textbf{g}}\Phi = (\text{det } g)^{-\frac 12}\partial_\mu((-\text{det }\textbf{g})^{\frac 12}\textbf{g}^{\mu\nu}\partial_\nu\Phi) = \nonumber\\&= \partial_v(2\partial_\rho\Phi + \lambda \Phi) - (\partial_\rho + \lambda)(\Gamma C\rho^2 A\partial_\rho\Phi + \rho K^i\partial_i\Phi) + \nabla_i(C^{-1}\partial^i\Phi - \rho K^i\partial_\rho \Phi).
\end{align}
We next integrate (\ref{wavecoord}) over $M(v,\rho)$. The final divergence term drops out, and in the middle term we integrate by parts to remove the derivatives $\partial_i$ acting on $\Phi$. Using $\mathcal{L}_{\partial_\rho}\text{vol}_g = \lambda\text{vol}_g$, we arrive at
\begin{equation} \label{intid}
    \partial_v I = \partial_\rho \int_{M(v,\rho)}\left(\Gamma C\rho^2A\partial_\rho \Phi -\rho (\nabla_i K^i)\Phi \right)\text{vol}_g.
\end{equation}
Since $K$ is divergence-free on the horizon, evaluating (\ref{intid}) on $\mathcal{H}$ yields
\begin{equation} \label{Ivr0}
    \partial_v I \:\overset{\mathcal{H}}{=}\:0.
\end{equation}
Hence $I = I_0$ is independent of $v$ on $\mathcal{H}$. To go further, we take a derivative of (\ref{intid}) and set $\rho = 0$,
\begin{equation}
    \partial_v \partial_\rho I \:\overset{\mathcal{H}}{=}\: 2\int_{M(v,0)}\left(A\partial_\rho\Phi - \partial_\rho(\nabla_i K^i)\Phi\right)\text{ vol}_g = 0. \label{Ivr1}
\end{equation}
The final equality holds since $A$ and $\partial_\rho(\nabla_i K^i)$ both vanish on $\mathcal{H}$. Observe that if $AI_0 \neq 0$ and $\Phi$ decays along $\mathcal{H}$ one instead finds that $\partial_v\partial_\rho I(v,0)$ approaches the constant $AI_0$, so that $\partial_\rho I(v,0)$ grows linearly in $v$. In the doubly degenerate case we must take a further derivative of (\ref{intid}),
    \begin{equation}
        \partial_v\partial_\rho^2I \:\overset{\mathcal{H}}{=}\:  \int_{M(v,0)}\left(6B \partial_\rho \Phi -3 \partial_\rho^2(\nabla_i K^i)\Phi\right)\text{vol}_g.
   \end{equation}
    If $\Phi \to 0$ as $v \to \infty$, the second term in the integrand decays as $v \to \infty$. The first term approaches $3BI_0$ provided $B$ is constant, implying the linear asymptotic growth of $\partial_\rho^2 I(v,0) \sim 3BI_0v$.
\end{proof}

We conclude this section by showing that the vanishing of $B$ corresponds to having a triply degenerate horizon in the sense that the function $\textbf{g}(V,V)$ considered in Section \ref{ddhsec} vanishes to cubic order in $\rho$ on $\mathcal{H}$. We again assume the conditions (\ref{a1}) and (\ref{a2}) are satisfied.  
The vector $V$ can be calculated in the spacetime using (\ref{Vdef}) and (\ref{ggnc}). Since each $m_I$ commutes with $k$ and $\partial_r$, it follows from (\ref{a2}) that $m_I$ is tangent to $M(v,r)$ for all $(v,r)$. We choose the extension of $\Gamma$ such that $\mathcal{L}_{m_I}\Gamma = 0$ holds everywhere. This is compatible with the choice in Lemma~\ref{lemel}, as can be seen by Lie-deriving (\ref{epde}) along a Killing vector $m_I$. The $m_I$ are then also tangent to $M(v,\rho)$. Writing $K_I = g(K,m_I) = m_I^\flat(K)$, we find
\begin{subequations}
\begin{align}
    V &= k - \rho h^{IJ}K_Im_J,\\
    \textbf{g}(V,V) &= \rho^2(\Gamma^2 F- h^{IJ}K_IK_J) = \rho^2\left(\Gamma A+ (g - h^{IJ}m_I^\flat \otimes m_J^\flat)(K,K)\right). \label{gvv}
\end{align}
\end{subequations}
As we saw in Section \ref{ddhsec}, the quadratic term in $\rho$ vanishes on $\mathcal{H}$ if $A = 0$ and $K$ is in the span of the $m_I$. Moreover, because $g - h^{IJ}m^\flat_Im^\flat_J$ is a non-negative bilinear form and $(g - h^{IJ}m_I^\flat \otimes m_J^\flat)(K,K)$ vanishes at $\rho = 0$, its first derivative in $\rho$ also vanishes. Using $A = B\rho + O(\rho^2)$, to cubic order in $\rho$ we obtain
\begin{equation} \label{V3}
    \textbf{g}(V,V) = \Gamma B\rho^3 + O(\rho^4).
\end{equation}
By the same reasoning as before we may interpret a horizon on which $A = B = 0$ as triply degenerate. The formula (\ref{V3}) provides a convenient method to calculate $B$ in practice, which we exploit in Appendix \ref{Aa}. It also shows explicitly that, at least for a wide class of spacetimes, $B$ does not depend on any coordinate choices made above. 

\section{Forms and uncharged scalars} \label{fsec}

As an example of a class of matter models for which the intrinsic rigidity theorem holds, we consider a generalisation of the theory in \cite{KLR} that contains (cosmological) Einstein-Maxwell(-Chern-Simons) theory, as well as many supergravity theories and their dimensional reductions. The theory is $(n+2)$-dimensional and has action 
\be \label{s1}
    S = \int_\mathcal{M} \left(\mathcal{R} - \frac 12f_{AB}(\Phi)\nabla_\mu\Phi^A\nabla^\mu \Phi^B - V(\Phi) - \sum_{p\geq 2} \frac{2}{p!}h_{IJ}^p(\Phi)\mathcal{F}_{\mu_1\dots\mu_p}^I\mathcal{F}^{J\mu_1\dots\mu_p}\right)\text{vol}_{\textbf{g}} + S_{\text{top}}. 
\ee
The matter content consists of uncharged scalars $\Phi^A$ and closed $p$-forms $\mathcal{F}_{\mu_1\dots\mu_p}^I$, where $A,B, \dots$ and $I,J,\dots$ are labels and $p$ ranges over $2\leq p \leq 1 + \lfloor \frac n2 \rfloor$. The functions $f_{AB}, h^p_{IJ}, V$ depend on the scalars, and $S_{\text{top}}$ is a topological term (not contributing to the energy-momentum tensor). The specific form of $S_{\text{top}}$ does not matter for the arguments leading to Theorem \ref{thm1} and Theorem \ref{thm2}. However, we are only able to prove that the individual matter fields on the horizon are preserved by the Killing vector $K$ in Theorem \ref{thm1} if $S_{\text{top}}$ satisfies the condition (\ref{ass1}) in Section \ref{secinh1}. This condition holds for a wide range of theories in which the topological term is of the form
\begin{equation} \label{OI}
    S_{\text{top}} = \int_{\mathcal{M}}\left(\sum_k \lambda_{J_0J_1\dots J_{k}}\mathcal{A}^{J_0} \wedge \mathcal{F}^{J_1}\wedge \dots \wedge \mathcal{F}^{J_k} + \sum_l\sigma(\Phi)_{J_0\dots J_l}  \mathcal{F}^{J_0}\wedge \dots\wedge \mathcal{F}^{J_l}\right).
\end{equation}
Here $\lambda_{J_0\dots J_{k}}$ are constants, $\mathcal{A}^I$ is a potential for $\mathcal{F}^I$ and the $\sigma(\Phi)_{J_0\dots J_l}$ are functions depending on the scalars. For example, this includes the bosonic part of 11D supergravity and 5D Einstein-Maxwell-Chern-Simons theory.

We will assume $f_{AB}$ and $h_{IJ}^p$ are positive definite, but the potential $V$ may have any sign. For simplicity, we assume the spacetime $\mathcal{M}$ is orientable. Varying the action with respect to $\textbf{g}$ leads to
\begin{align} \label{bigemt}
    \mathcal{T}_{\mu\nu} =\: \frac 14f_{AB}&\left(2\nabla_\mu\Phi^A\nabla_\nu\Phi^B -\nabla_\rho\Phi^A\nabla^\rho \Phi^B\textbf{g}_{\mu\nu}\right) - \frac 12V\textbf{g}_{\mu\nu}\nonumber\\ &+ \sum_{p\geq 2} \frac{2}{(p-1)!}h_{IJ}^p\left(\mathcal{F}^{I}_{\mu\rho_1\dots\rho_{p-1}}\mathcal{F}_\nu^{J\rho_1\dots\rho_{p-1
    }} - \frac{1}{2p}\mathcal{F}_{\mu_1\dots\mu_p}^I\mathcal{F}^{J\mu_1\dots\mu_p}\textbf{g}_{\mu\nu}
    \right).
\end{align}
Here we have suppressed the $\Phi$-dependence of $f_{AB}, V$ and $h_{IJ}^p$. The tensor (\ref{bigemt}) satisfies the NEC and also the SEC if $V \leq 0$. The equations of motion are
\begin{subequations} \label{mateqs}
\begin{align}
    &\text{d}\star_{\textbf{g}} (h_{IJ}^p\mathcal{F}^J) + \mathcal{O}_I(\Phi,\mathcal{F}) = 0, \label{matF}\\
    &{^\textbf{g}}\nabla_\mu(f_{AB}\nabla^\mu \Phi^B) -\tfrac12 f_{BC,A}\nabla_\mu\Phi^B\nabla^\mu\Phi^C - V_{,C} \nonumber\\&\hspace{2.7cm}- \sum_{p\geq2}\tfrac{2}{p!}h_{IJ,A}^p\mathcal{F}_{\mu_1\dots\mu_p}^I\mathcal{F}^{J\mu_1\dots\mu_p} + \mathcal{Q}_A(\Phi,\mathcal{F}) = 0. \label{matphi}
\end{align}
\end{subequations}
The terms $\mathcal{O}_I$ and $\mathcal{Q}_A$ represent contributions from the topological term in the action, and the comma denotes a derivative with respect to a scalar field (e.g. $V_{,C} = \partial V/ \partial\Phi^C$). The matter equation for the scalar fields (\ref{matphi}) is not needed in the arguments below and is only included for completeness.
\subsection{Horizon data} Suppose $\mathcal{H}$ is an extremal horizon in this theory with generator $k$ and compact $n$-dimensional cross-section $M$. The matter fields $\Phi^A$ and $\mathcal{F}^I$ are assumed to be preserved by $k$. In addition to the data $(g,X,T,U)$ defined in Section \ref{ehors}, the horizon data consists of induced matter fields on $M$. Each scalar $\Phi^A$ may be pulled back to a scalar $\phi^A$ on $M$, and similarly each $p$-form $\mathcal{F}^I$ induces a closed $p$-form $B^I$. Moreover, every $\mathcal{F}^I$ defines a $(p-2)$-form $C^I$ on $M$ via
\begin{equation*}
    \iota_k \mathcal{F} ^I
    \overset{\mathcal{H}}{=}\: k \wedge C^I.
\end{equation*}
The existence of $C^I$ follows from the fact that $\mathcal{T}(k,k)$ vanishes on $\mathcal{H}$. Indeed, the norm of $\iota_k \mathcal{F}^I$ must be non-negative on $\mathcal{H}$ since $\iota_k\mathcal{F}^I$ is orthogonal to the null vector $k$. Hence, choosing a basis at a point such that $h^p_{IJ}$ is diagonal, from $\mathcal{T}(k,k) = 0$ we find that $\iota_k \mathcal{F}^I$ is null on $\mathcal{H}$. This can only happen if $k \wedge \iota_k \mathcal{F}^I = 0$, proving $C^I$ is well-defined on $\mathcal{H}$. Since $\iota_kC^I = 0$ we may view $C^I$ as a $(p-2)$-form on $M$. 

The data $(T,U)$ can be expressed in terms of the induced matter fields $(\phi^A, B^I,C^I)$ as
\begin{subequations} \label{id}
\begin{align}
    T_{ab} &= \frac 14f_{AB}\left(2\nabla_a \phi^A\nabla_b\phi^B - \langle \text{d}\phi^A, \text{d}\phi^B\rangle g_{ab}\right) - \frac 12 Vg_{ab} - \sum_{p\geq 3} \frac{2}{(p-3)!}h_{IJ}^pC^I_{ac_1\dots c_{p-3}}C^{Jc_1\dots c_{p-3}}_b\nonumber\\ &\hspace{.5cm}+ \sum_{p\geq 2} h_{IJ}^p\left(\frac{2}{(p-1)!}B^{I}_{ac_1\dots c_{p-1}}B_b^{Jc_1\dots c_{p-1
    }} + \left(\langle C^I, C^J\rangle - \langle B^I, B^J\rangle\right) g_{ab}
    \right), \\
    U &= -\frac 14f_{AB}\langle\text{d}\phi^A, \text{d}\phi^B\rangle  - \frac 12 V - \sum_{p \geq 2} h_{IJ}^p\left( \langle C^I, C^J\rangle + \langle B^I, B^J\rangle \right).
\end{align}
\end{subequations}
Here we write 
\begin{equation*}
    \langle B^I, B^J\rangle = \frac{1}{p!}B^I_{a_1\dots a_{p}}B^{Ja_1\dots a_{p}}
\end{equation*}
for the $g$-inner product on forms, and similarly for $\langle C^I, C^J \rangle$ and $\langle \text{d}\phi^A, \text{d}\phi^B\rangle$. We next compute the 1-form $\beta$ and the function $\alpha$ from their definitions (\ref{betadef}, \ref{alphadef}). A convenient way to do this is to express the energy-momentum tensor in Gaussian null coordinates and identify $\alpha$ and $\beta$ with the leading order terms of certain components of (\ref{bigemt}) as explained in Section \ref{nhgs}. The data $(\phi^A,B^I,C^I)$ can similarly be viewed as components of $(\Phi^A, \mathcal{F}^I)$ on $\mathcal{H}$ in Gaussian null coordinates,
\begin{equation*}
    \phi^A = \Phi^A\vert_{r = 0}, \hspace{.6cm} B^I = \tfrac{1}{p!}\mathcal{F}^I_{i_1\dots i_p}\vert_{r = 0}\:\text{d}x^{i_1}\wedge\dots\wedge\text{d}x^{i_p}, \hspace{.6cm} C^I = \tfrac{1}{(p-2)!}\mathcal{F}^I_{vri_1\dots i_{p-2}}\vert_{r = 0}\:\text{d}x^{i_1}\wedge\dots\wedge \text{d}x^{i_{p-2}}.
\end{equation*}
In this way we find
\begin{subequations}
\begin{align}
    \iota_Y \beta &=  \sum_{p\geq2}2h^p_{IJ}\left(\left\langle \iota_Y B^I, \text{d}C^J - X \wedge C^J\right\rangle-\left\langle C^I, \iota_Y(\text{d}C^J - X \wedge C^J)\right\rangle \right), \label{betaex1} \\
    \alpha &= \sum_{p \geq 2} 2h_{IJ}^p\left\langle \text{d}C^I - X \wedge C^ I, \text{d}C^J - X \wedge C^ J\right\rangle. \label{alphaex1}
\end{align}
\end{subequations}
Here $Y$ is an arbitrary vector field on $M$ and we used the fact that $\mathcal{F}^I$ is closed. Note that the function $r^2\alpha = \mathcal{T}_\text{H}(e_+,e_+)$ is non-negative as required by the null energy condition. It is now straightforward to decompose $X$ into $K$ and $\Gamma$ and compute $\gamma$ from (\ref{Ldef}),
\begin{equation} \label{L}
    \gamma = \frac 12 f_{AB}\mathcal{L}_K\phi^A\mathcal{L}_K\phi^B + \sum_{p\geq 2} 2h_{IJ}^p\left\langle \iota_K B^I - \text{d}(\Gamma C^I), \iota_K B^J - \text{d}(\Gamma C^J)\right\rangle.
\end{equation}
As anticipated, this expression is non-negative. Integrating the divergence identity (\ref{magid}) shows that $K$ either vanishes or is a Killing vector, and $\gamma = 0$. The vanishing of $\gamma$ is equivalent to 
\begin{equation} \label{easymatter}
    \mathcal{L}_K \phi^A = 0, \hspace{.8cm} \iota_K B^I = \text{d}(\Gamma C^I).
\end{equation}
These conditions together with $\mathcal{L}_K g = 0$ ensure that (\ref{simp2}) holds. The constant $A$ in (\ref{Aclean}) becomes

\begin{equation} 
    A = -\frac{\vert K \vert^2}{2\Gamma} + \frac 12\Delta \Gamma + \frac 1n\Gamma V - \frac 2n\Gamma \sum_{p \geq 2}h_{IJ}^p \left((n+1-p)\langle C^I, C^J\rangle + (p-1) \langle B^I,  B^J\rangle\right).
\end{equation}
Note that the matter terms are non-positive for $V \leq 0$, when the strong energy condition is satisfied. 

In addition to the above, the fields $(\phi^A,B^I,C^I)$ satisfy equations of motion coming from the matter equations (\ref{mateqs}). The topological term $\mathcal{O}_I$ induces a $(n+3-p)$-form $O_I$ on $M$ by restriction, as well as a $(n+1-p)$-form $P_I$ via
\begin{equation} \label{defP}
    \iota_k\mathcal{O}_I \:\overset{\mathcal{H}}{=}\:k\wedge P_I.
\end{equation}
The existence of $P_I$ can be deduced from (\ref{matF}) as follows. Each $\mathcal{F}^I$ can be written in the basis (\ref{onbs}) as a sum of terms of the form
\begin{align*}
    \textbf{e}^+ \wedge \textbf{e}^- \wedge \hat{\textbf{e}}^{i_1} \wedge \dots \wedge \hat{\textbf{e}}^{i_{p-2}} , \hspace{.6cm} \textbf{e}^- \wedge \hat{\textbf{e}}^{i_1} \wedge \dots \wedge \hat{\textbf{e}}^{i_{p-1}}, \hspace{.6cm} \textbf{e}^+ \wedge \hat{\textbf{e}}^{i_1} \wedge \dots \wedge \hat{\textbf{e}}^{i_{p-1}}, \hspace{.6cm}   \hat{\textbf{e}}^{i_1} \wedge \dots \wedge \hat{\textbf{e}}^{i_{p}}. 
\end{align*}
The coefficient of the third term must be $O(r)$ as  $k \wedge \iota_k \mathcal{F}^I$ vanishes on $\mathcal{H}$. By (\ref{matF}), we have $\iota_k \mathcal{O}_I = \text{d}\iota_k \star_{\textbf{g}} (h_{IJ}^p\mathcal{F}^J)$. Only the final two types of terms contribute to this expression on the horizon since $k =\textbf{e}_+ + O(r^2)$. Both of these contributions are wedge products of $k$ with a $(n+1-p)$-form, so that we can write $\iota_k\mathcal{O}_I$ as in (\ref{defP}).

The other topological term $\mathcal{Q}_A$ induces a function $Q_A$. The equations of motion on $M$ may be obtained by a tedious calculation in the basis (\ref{onbs}). They can be further simplified using (\ref{easymatter}), resulting in
\begin{subequations} \label{hormatF}
\begin{align}
    &\text{d}\star (\Gamma h^p_{IJ}B^J) + \iota_K \star h^p_{IJ}C^J + \Gamma P_I = 0, \label{mF1}\\
    &\text{d}\star h^p_{IJ}C^J - O_I = 0, \label{mF2} \\ 
    &\nabla_a(\Gamma f_{AB} \nabla^a \phi^B) - \tfrac 12\Gamma f_{BC,A}\langle \text{d}\phi^B,\text{d}\phi^C\rangle - \Gamma V_{,A} \nonumber \\ &\hspace{2.7cm} - \sum_{p\geq 2} 2\Gamma h^p_{IJ,A}\left(\langle B^I, B^J\rangle -\langle C^I, C^J \rangle\right) + \Gamma Q_A = 0.\label{phieom}
\end{align}
\end{subequations}
The hodge star\footnote{We choose an orientation vol$_g$ on $M$ so that $\iota_k\text{vol}_{\textbf{g}} = k \wedge \text{vol}_g$ holds on $\mathcal{H}$. Note that $M$ is orientable if $\mathcal{M}$ is.} is taken with respect to $g$ and a comma denotes a derivative with respect to $\phi^A$.

Using the horizon matter data we can define matter fields in the near-horizon geometry
\begin{equation} \label{nhfields}
    \Phi^A_{\text{H}} = \phi^A, \hspace{1cm} \mathcal{F}^I_{\text{H}} = -\text{d}(r\text{d}v \wedge C^I) + B^I.
\end{equation}
These are such that the matter data induced by (\ref{nhfields}) returns $(\phi^A,B^I, C^I)$, and the matter equations (\ref{mateqs}) for the near-horizon geometry are equivalent to (\ref{hormatF}). One may also think of (\ref{nhfields}) as the leading order approximation or near-horizon limit of the spacetime matter fields  away from $\mathcal{H}$.
\subsection{Inheritance of symmetry} \label{secinh1}
The conditions (\ref{easymatter}) coming from the divergence identity imply that $K$ preserves $\phi^A$ and $B^I$, but for $p > 2$ showing the invariance of $C^I$ is less straightforward. We are able to show $K$ preserves $C^I$ only under an assumption on the topological term in the action, which is that 
\begin{equation} \label{ass1}
    \iota_K O_I = \text{d}(\Gamma P_I).
\end{equation}
For $p=2$ we have $O_I = 0$ for dimensional reasons and we will see (\ref{ass1}) is always satisfied. Moreover, if the topological term is of the form (\ref{OI}), we have 
\begin{equation} \label{OI2}
    \mathcal{O}_I(\Phi, \mathcal{F}) = -\frac 14\sum_k (k+1)\lambda_{IJ_1\dots J_{k}}\mathcal{F}^{J_1}\wedge \dots \wedge \mathcal{F}^{J_k} -\frac{1}{4} \sum_l(l+1)\text{d}(\sigma(\Phi)_{IJ_1\dots J_l}) \wedge \mathcal{F}^{J_1}\wedge \dots\wedge \mathcal{F}^{J_l},
\end{equation}
so that (\ref{ass1}) holds as a consequence of (\ref{easymatter})\footnote{We take $\lambda_{J_0\dots J_k}$ and $\sigma(\Phi)_{J_0\dots J_k}$ to be graded-symmetric in the indices labelling the field strengths, i.e. swapping the indices $J_i$ and $J_j$ introduces a sign $(-1)^{p_ip_j}$, where $p_i$ and $p_j$ denote the degree of $\mathcal{F}^{J_i}$ and $\mathcal{F}^{J_j}$ respectively.}.  Under the condition (\ref{ass1}), $K$ preserves all the horizon data and the near-horizon matter fields inherit the symmetries of the near-horizon geometry.
\begin{prop} \label{propm1}
    Consider an extremal horizon in the theory \eqref{s1} with matter data $(\phi^A,B^I,C^I)$ on a compact cross-section $M$. Suppose the topological term satisfies the condition \eqref{ass1}. 
    \begin{enumerate}[(i)]
        \item If the horizon data is rotating, the Killing vector $K$ in Theorem~\ref{thm1} preserves $(\phi^A,B^I,C^I)$. \vspace{.1cm}
        \item The near-horizon matter fields \eqref{nhfields} are preserved by the Killing vectors generating the isometries in Theorem~\ref{thm2} and, if the horizon data is both static and rotating, by the Killing vectors in Proposition~\ref{propads3}.
    \end{enumerate}
\end{prop}
\begin{proof}
    $(i)$ The invariance of $\phi^A$ and $B^I$ follows immediately from (\ref{easymatter}) using $\mathcal{L}_K = \text{d}\iota_K + \iota_K\text{d}$ and the fact that $B^I$ is closed. We also find $\iota_K\text{d}(\Gamma C^I) = 0$, which for $p = 2$ implies $\mathcal{L}_K C^I = 0$ since $C^I$ is a $0$-form and $\mathcal{L}_K\Gamma = 0$. For $p > 2$ we make use of the matter equations (\ref{hormatF}). Taking the exterior derivative of (\ref{mF1}) and hooking $K$ into (\ref{mF2}), we obtain
    \begin{equation*}
        \mathcal{L}_K\star h^p_{IJ}C^J =  \text{d}\iota_K\star h^p_{IJ}C^J + \iota_K\text{d}\star h^p_{IJ}C^J  = \iota_K O_I-\text{d}(\Gamma P_I).
    \end{equation*}
    If (\ref{ass1}) holds we deduce $h^p_{IJ}\mathcal{L}_KC^J = 0$ and hence $\mathcal{L}_K C^I = 0$ because $h^p_{IJ}$ is non-degenerate. Conversely, the same computation shows that (\ref{ass1}) must always hold when $p = 2$.

    $(ii)$ The fields (\ref{nhfields}) are invariant under $\partial_v$ and $v\partial_v - r\partial_r$ by construction, and they are invariant under $K$ by the arguments above. In addition, it is straightforward to verify that $m$ defined in (\ref{mkv}) preserves the near-horizon fields as a consequence of (\ref{easymatter}) for any value of the constant $A$. In the case where the horizon is both static and rotating, it was shown in Section \ref{ads3} that the function $\alpha$ in (\ref{alphaex1}) vanishes. This happens if and only if
\begin{equation*}
    \text{d}(\Gamma C^I) = K^\flat \wedge C^I.
\end{equation*}
Together with (\ref{easymatter}) we deduce $\iota_K (K^\flat \wedge C^I) = 0$ and so $\vert K \vert^2 C^I = K^\flat \wedge \hat{C}^I$ for a $(p-3)$-form $\hat{C}^I$ (if $p = 2$ then $C^I = 0$). We use a hat to denote forms on the orbit space $N$ of $K$. We also find $\iota_K B^I = 0$ and $\text{d}(\Gamma \hat{C}^I) = 0$ using d$(\Gamma^{-1}K^\flat) = 0$. In local coordinates where $K = \partial_{\chi}$, the matter fields become
\begin{equation} \label{matads3}
    \Phi^A_{\text{H}} = \hat{\phi}^A, \hspace{1cm}\mathcal{F}_{\text{H}}^I = \hat{\Gamma}(\text{d}v\wedge \text{d}\rho \wedge\text{d}\chi \wedge \hat{C}^I) + \hat{B}^I.
\end{equation}
The form $\text{d}v \wedge \text{d}\rho\wedge \text{d}\chi$ is a constant multiple of the volume form of the AdS$_3$ factor in (\ref{gads3}), which implies that the fields (\ref{matads3}) are preserved by any of the Killing vectors in Proposition \ref{propads3}.
\end{proof} 

Similar arguments as in the proof of Theorem \ref{thm2} may be applied to show one can introduce matter fields $ \overline{\Phi}^A_{\text{H}}$ and $\overline{\mathcal{F}}_{\text{H}}^I$ in the extended near-horizon geometry $\Sigma \times M$ that are invariant under the orientation-preserving isometries of AdS$_2$, 2D Minkowski space or dS$_2$. For the rotating case, in terms of local coordinates $(y^i, \chi)$ on $M$ such that $K = \partial_\chi$ we have $\Phi^A_{\text{H}} = \phi^A(y)$ and
\begin{equation*}
\mathcal{F}_{\text{H}}^I = \Gamma \text{d}v \wedge \text{d}\rho \wedge C^I + \tfrac{1}{p!}B^I_{y^{i_1}\dots y^{i_p}}\text{d}y^{i_1} \wedge \dots \wedge  \text{d}y^{i_p} + \tfrac{1}{(p-1)!}B^I_{\chi y^{i_1}\dots y^{i_{p-1}}}(\text{d}\chi + \rho\text{d}v)\wedge \text{d}y^{i_1}\wedge \dots \wedge \text{d}y^{i_{p-1}}.
\end{equation*}
Here $\Gamma$ and the components of $B^I,C^I$ are functions of $y$ only. The transformation to global coordinates on $\Sigma$ corresponds to replacing $\text{d}v \wedge \text{d}\rho \mapsto \text{d}\tau \wedge \text{d}\sigma$ and $\text{d}\chi + \rho\text{d}v \mapsto \text{d}\chi + \sigma\text{d}\tau$. In the AdS$_3$ case we can replace the form $\text{d}v \wedge \text{d}\rho \wedge \text{d}\chi$ in (\ref{matads3}) by (a constant multiple of) the volume form of the full AdS$_3$ spacetime to obtain matter fields invariant under $SO(2,2)$.

\section{Yang-Mills fields and charged matter} \label{gsec} 
For our second class of examples we consider a gauge field coupled to charged matter. Special cases of interest include Einstein-Yang-Mills(-Chern-Simons) theory, as well as the Einstein-Maxwell-charged scalar field model. Near-horizon geometries in these theories have been studied previously in four spacetime dimensions in \cite{LL13} and \cite{LL16} respectively.

Let $G$ be a compact Lie group with Lie algebra $\mathfrak{g}$. For simplicity, we will assume $G$ is a matrix Lie group. The field content consists of a connection on a principal $G$-bundle $\mathcal{P}$ over an $(n+2)$-dimensional spacetime $\mathcal{M}$, locally represented by a $\mathfrak{g}$-valued 1-form $\mathcal{A}$. The corresponding field strength is $\mathcal{F} = \text{d}\mathcal{A} + \frac 12[\mathcal{A},\mathcal{A}]$. We also include charged fields $\Phi^I$, which are sections of vector bundles over $\mathcal{M}$ associated to (real or complex) representations of $G$. The action reads
\be \label{ymaction} 
    S = \int_{\mathcal{M}} \left(\mathcal{R} - \frac 12 f_{IJ}(\Phi)\langle \mathcal{D}_\mu \Phi^I, \mathcal{D}^\mu \Phi^J\rangle - V(\Phi) + h(\Phi)\text{Tr}(\mathcal{F}_{\mu\nu}\mathcal{F}^{\mu\nu})\right)\text{vol}_{\textbf{g}} + S_{\text{top}}. 
\ee
Here $\mathcal{D}\Phi^I = \text{d}\Phi^I + \mathcal{A}\cdot \Phi^I$ is the covariant derivative of $\Phi^I$, where $\cdot$ denotes the action of $\mathcal{A}$ on $\Phi^I$. The bracket $\langle \cdot , \cdot \rangle$ denotes a $G$-invariant Euclidean or Hermitian inner product on a representation space\footnote{For (\ref{ymaction}) to be well-defined, we require $f_{IJ} = 0$ unless $\Phi^I$ and $\Phi^J$ are sections of the same bundle.}. To ensure the NEC holds, we assume the inner product $-h\text{Tr}(\cdot,\cdot)$ on $\mathfrak{g}$ and the matrix $f_{IJ} = \overline{f}_{JI}$ are positive definite. As before, the topological term needs to satisfy a condition in order for the horizon matter fields to inherit the symmetry, see equation (\ref{ass2}) in Section \ref{62}. The allowed terms include
\begin{equation} \label{stop}
    S_{\text{top}} = \int_{\mathcal{M}}
    \lambda \omega_{2k+1} \hspace{.5cm}(\text{for } n = 2k-1), \hspace{.8cm} S_{\text{top}} = \int_{\mathcal{M}} \sigma(\Phi)\text{Tr}(\mathcal{F}^{\wedge (k+1)}) \hspace{.5cm}(\text{for } n = 2k).
\end{equation}
Here $\omega_{2k+1}$ is the Chern-Simons form satisfying d$\omega_{2k+1} = \text{Tr}(\mathcal{F}^{\wedge (k+1)})$, $\lambda$ is a constant and $\sigma$ is a function depending on the fields $\Phi^I$. In particular, the condition is satisfied for Einstein-Yang-Mills-Chern-Simons theory.

Under a ($G$-valued) gauge transformation $\tau$, we have 
\begin{equation*}
    \mathcal{A} \mapsto \tau \mathcal{A}\tau^{-1} - (\text{d}\tau )\tau^{-1}, \hspace{.5cm} \mathcal{F} \mapsto \tau \mathcal{F}\tau^{-1}, \hspace{.5cm} \Phi^I \mapsto \tau \cdot \Phi^I.
\end{equation*}
This transformation preserves the action (\ref{ymaction}), provided $f_{IJ}, h$ and the potential $V$ are invariant. Note that $\cdot$ is the trivial action if $\Phi^I$ is uncharged. We are interested in configurations invariant under a Killing vector $k$, by which we mean that in any gauge there exists a $\mathfrak{g}$-valued function $\sigma_k$ such that 
\begin{equation} \label{kinv}
    \mathcal{L}_k \mathcal{A} = \mathcal{D}\sigma_k = \text{d}\sigma_k + [\mathcal{A},\sigma_k], \hspace{.8cm} \mathcal{L}_k\mathcal{F} = [\mathcal{F},\sigma_k], \hspace{.8cm} \mathcal{L}_k \Phi^I = -\sigma_k \cdot \Phi^I.
\end{equation}
The energy-momentum tensor is a gauge-covariant version of (\ref{bigemt})
\begin{equation} \label{bigemt2}
    \mathcal{T}_{\mu\nu} = \frac 14 f_{IJ}\left(2\langle \mathcal{D}_{(\mu}\Phi^I,\mathcal{D}_{\nu)}\Phi^J\rangle - \langle\mathcal{D}_\rho\Phi^I,\mathcal{D}^\rho\Phi^J\rangle \textbf{g}_{\mu\nu}\right) - \frac 12V\textbf{g}_{\mu\nu} - 2h\text{Tr}\left(\mathcal{F}_{\mu\rho}\mathcal{F}_\nu^{\:\rho} - \tfrac 14\mathcal{F}_{\rho\sigma}\mathcal{F}^{\rho\sigma}\textbf{g}_{\mu\nu}\right).
\end{equation}
To write down the equations of motion, we introduce orthonormal bases $t_i$ of $\mathfrak{g}$ (with respect to $-\text{Tr}(\cdot,\cdot)$) and $e_A$ of the representation spaces (with respect to $\langle \cdot ,\cdot\rangle$). The equations are
\begin{subequations}\label{DFeom}
\begin{align} 
    &\mathcal{D}^\mu(h\mathcal{F}_{\mu\nu}^i) - \tfrac 18\left(f_{IJ}c^i_{AB}\overline{\mathcal{D}_\nu \Phi^{IA}}\Phi^{JB} + \text{c.c.}\right) + \mathcal{O}^i_\nu(\mathcal{A},\Phi) = 0, \label{nafeom}\\
    &\mathcal{D}_\mu(f_{IJ}\mathcal{D}^\mu\Phi^J) -  f_{MN,IJ}\Phi^J\langle \mathcal{D}_\mu \Phi^M, \mathcal{D}^\mu\Phi^N\rangle - 2V_{,IJ}\Phi^J + 2h_{,IJ}\Phi^J\text{Tr}\left(\mathcal{F}_{\mu\nu}\mathcal{F}^{\mu\nu}\right) + \mathcal{Q}_I = 0. \label{phiceomb}
\end{align}
\end{subequations}
Here we expanded $\mathcal{D}^\mu\mathcal{F}_{\mu\nu} = (\mathcal{D}^\mu\mathcal{F}_{\mu\nu}^i)t_i$ and $\mathcal{D}_\nu\Phi^I = (\mathcal{D}_\nu\Phi^{IA})e_A$. The $c^i_{AB}$ are structure constants defined by $t_i \cdot e_B = \sum_A c_{AB}^ie_A$. ``c.c." stands for complex conjugate and $\mathcal{O}, \mathcal{Q}_I$ represent contributions from $S_{\text{top}}$. In (\ref{phiceomb}) we assumed the functions $f_{MN},V$ and $h$ only depend on $\Phi$ through the inner product $x^{IJ} = \langle \Phi^I,\Phi^J\rangle$, and the comma denotes the partial derivative with respect to $x^{IJ}$ (note that a single derivative is being taken). This assumption is only for simplicity: just like in the uncharged case, the equation (\ref{phiceomb}) is not needed to prove the inheritance of symmetry in Proposition \ref{propym}.

\subsection{Horizon data} We follow the approach in Section \ref{fsec} to study extremal horizons in this theory. As will become clear, some of the arguments are more subtle if the gauge field is non-abelian or the matter is charged.

Given an extremal horizon $\mathcal{H}$ generated by $k$ with $n$-dimensional cross-section $M$, there is an induced connection $A$ with curvature $B$ and covariant derivative $D$ on the bundle $P = i^*\mathcal{P}$ obtained by pulling back $\mathcal{A}$ along the inclusion $i: M \to \mathcal{M}$. We can similarly pull back $\Phi^I$ to obtain fields $\phi^I$ on $M$. The condition $\mathcal{T}(k,k) = 0$ on $\mathcal{H}$ implies 
\begin{equation} \label{tkk}
    \mathcal{D}_k\Phi^I \:\overset{\mathcal{H}}{=}\:0, \hspace{1.5cm} k \wedge \iota_k\mathcal{F} \:\overset{\mathcal{H}}{=}\:0.
\end{equation} Here $\mathcal{D}_k \Phi^I = \iota_k \mathcal{D}\Phi^I$. It follows that we can define a section $C$ of the adjoint bundle Ad ~$P$ by 
\begin{equation*}
    \iota_k \mathcal{F} \:\overset{\mathcal{H}}{=}\: C k.
\end{equation*}
The energy-momentum data $(T,U)$ is easily computed from (\ref{bigemt2})
\begin{subequations}
\begin{align}
    T_{ab} &= \frac 14f_{IJ}\left(2\langle D_{(a}\phi^I, D_{b)}\phi^J\rangle - \langle D_c\phi^I, D^c\phi^J\rangle g_{ab}\right)
   - \frac 12 Vg_{ab} \nonumber \\&\hspace{4.3cm} - h\text{Tr}\left(2B_{ac}B_b^{\:\:c} + \tfrac 12g_{ab}(2C^2 - B_{cd}B^{cd})\right), \\
    U &= - \frac 14f_{IJ}\langle D_a\phi^I, D^a\phi^J\rangle - \frac 12 V +\frac 12 h\text{Tr}\left(2C^2 + B_{ab}B^{ab}\right).
\end{align}
\end{subequations}
Unlike in the theory (\ref{s1}), there are contributions to $\alpha$ and $\beta$ that can a priori not be expressed in terms of the data $(\phi^I,A,B,C)$. These involve fields $\psi^I$ and an Ad-valued 1-form $H$ on $M$, defined by
\begin{equation} \label{Hdef}
    \mathcal{D}(\mathcal{D}_k \Phi^I) \:\overset{\mathcal{H}}{=}\: \psi^I \:k, \hspace{.5cm} \iota_Y \mathcal{F} \:\overset{\mathcal{H}}{=}\: \iota_YB + \iota_Y H \:k. 
\end{equation}
Here $Y$ is an arbitrary section of $TM$. It follows from (\ref{tkk}) that $\psi^I$ and $H$ are well-defined. Equivalently, in Gaussian null coordinates
\begin{equation*}
    \psi^I = \partial_r(\mathcal{D}_v\Phi^I)\vert_{r = 0}, \hspace{1cm} H = \mathcal{F}_{ar}\vert_{r = 0}\:\text{d}x^a.
\end{equation*}
The fields $\psi^I$ vanish in the uncharged case due to (\ref{kinv}), and in the abelian case $H$ constitutes data extrinsic to the horizon in the sense that the components $\mathcal{F}_{ar}$ decouple from the near-horizon limit. In general however there are contributions to the horizon matter equations coming from $\psi^I$ and the exterior covariant derivative\footnote{In order to apply $\mathcal{D}_k$ to an object defined on $M$, we extend it to $\mathcal{M}$ in any way such that it is preserved by $k$ in the sense of (\ref{kinv}). In this case we find $\mathcal{D}_k H = \mathcal{L}_kH + [\iota_k \mathcal{A},H] = [\iota_k \mathcal{A}-\sigma_k,H]$ on $M$. Note that $\mathcal{D}_k H$ is algebraic in $H$ and hence its value on $M$ is independent of the extension ($H$ is in fact already naturally defined on $\mathcal{H}$ by (\ref{Hdef})).} $\mathcal{D}_k H$ along $k$ (but not $H$ itself), which should be considered data intrinsic to $\mathcal{H}$. 
The  expressions for $\alpha$ and $\beta$ are 
\begin{subequations}\label{abym}
\begin{align} 
    \beta_a &= \frac 14f_{IJ}(\langle \psi^I, D_a\phi^J\rangle + \langle D_a\phi^I,\psi^J\rangle)- 2h\text{Tr}\left((D^bC - \mathcal{D}_kH^b - X^bC)(B_{ab} - g_{ab}C)\right), \\
    \alpha &= \frac 12f_{IJ}\langle \psi^I, \psi^J\rangle - 2h\text{Tr}\left(\vert DC - \mathcal{D}_kH - XC\vert^2\right). \label{alphaex2}
\end{align}
\end{subequations}
Here $\vert \cdot \vert^2$ denotes the $g$-norm on (matrix-valued) 1-forms, e.g. $\vert H \vert^2 = H_aH^a$. To derive (\ref{abym}) we calculated the relevant components of $\mathcal{T}$ in the basis (\ref{onbs}) and used the fact that $\mathcal{D}\mathcal{F} = 0$. The function $\gamma$ in (\ref{Ldef}) is computed to be
\begin{equation} \label{gamma2}
    \gamma = \tfrac 12f_{IJ}\langle D_K\phi^I - \Gamma \psi^I,D_K\phi^J - \Gamma \psi^J\rangle - 2h\text{Tr}\left( \vert \iota_K B - D(\Gamma C) + \Gamma \mathcal{D}_kH\vert^2\right).
\end{equation}
Just like in (\ref{L}), the function $\gamma$ is a sum of non-negative terms. Recall that the proof of Theorem~\ref{thm1} implies that $\gamma$ must vanish on compact $M$. Therefore 
\begin{equation} \label{fint}
    D_K\phi^I = \Gamma \psi^I, \hspace{.8cm} \iota_K B = D(\Gamma C) - \Gamma\mathcal{D}_k H.
\end{equation}
We will show that, under a mild condition on the topological term in (\ref{ymaction}), the terms involving $\psi^I$ and $\mathcal{D}_kH$ can be eliminated from (\ref{fint}) (in fact $\mathcal{D}_kH$ vanishes but $\psi^I$ does not in general). The resulting equations allow us to deduce the intrinsic data $(\phi^I,A,B,C)$ is invariant under $K$. Moreover, the matter equations induced by (\ref{DFeom}) become equivalent to the equations of motion for near-horizon matter fields
\begin{equation} \label{nhfields2}
    \Phi^I_{\text{H}} = \phi^I, \hspace{1cm} \mathcal{A}_{\text{H}} = -C r\text{d}v + A, \hspace{1cm} \mathcal{F}_{\text{H}} = C \text{d}v\wedge\text{d}r - rDC\wedge\text{d}v + B.
\end{equation}
Note that we must include  $\mathcal{A}_{\text{H}}$ as a separate matter field because it does not just enter the equations through $\mathcal{F}_{\text{H}}$. The interpretation of (\ref{nhfields2}) as an approximation to the spacetime matter fields away from $\mathcal{H}$ is also discussed in Section \ref{62}.

\subsection{Inheritance of symmetry} \label{62}To show the inheritance of symmetry for the matter fields, we assume the topological term in (\ref{DFeom}) is such that
\begin{equation} \label{ass2}
    \iota_k \mathcal{O} \:\overset{\mathcal{H}}{=}\:0.
\end{equation}
This condition follows from (\ref{nafeom}) if the gauge field is abelian. It is also satisfied as a consequence of (\ref{tkk}) if $S_{\text{top}}$ is of the form (\ref{stop}), in which case
\begin{equation*}
    \mathcal{O}(\mathcal{A},\Phi) = \frac{k+1}{4}\lambda\star_{\textbf{g}}\mathcal{F}^{\wedge k} \hspace{.3cm}(\text{for } n = 2k-1), \hspace{.7cm} \mathcal{O}(\mathcal{A},\Phi) =\frac{k+1}{4}\star_\textbf{g} \left(\text{d}\left(\sigma (\Phi)\right)\wedge \mathcal{F}^{\wedge k}\right) \hspace{.3cm}(\text{for } n = 2k).
\end{equation*}
We will make use of the following result, which is based on \cite{LL13}. Note that we do not require $\mathfrak{g}$ to be semisimple.
\begin{lemma} \label{lem}
    Let $\Theta$ be a section of an associated vector bundle $E$ over $\mathcal{M}$, which is preserved by $k$ in the sense of \eqref{kinv}. Suppose $\mathcal{D}_k(\mathcal{D}_k\Theta)$ vanishes at a point $p \in \mathcal{M}$. Then $\mathcal{D}_k\Theta = 0$ at $p$.
\end{lemma}
\begin{proof}
    In any gauge we have $\mathcal{D}_k \Theta = \mathcal{L}_k\Theta + \iota_k\mathcal{A} \cdot \Theta = (\iota_k\mathcal{A} - \sigma_k)\cdot \Theta$ and $\mathcal{D}_k (\iota_k\mathcal{A} - \sigma_k) = 0$. Hence
    \begin{equation*}
    \mathcal{D}_k(\mathcal{D}_k\Theta) = (\iota_k\mathcal{A} - \sigma_k) \cdot \left((\iota_k\mathcal{A} - \sigma_k) \cdot \Theta\right).
    \end{equation*}
    Let $a \in \mathfrak{g}$ be the value of $\iota_k\mathcal{A} - \sigma_k$ at $p$. Since $G$ is compact, the map $E_p \to E_p, \: v \mapsto a \cdot v$ is diagonalisable over $\mathbb{C}$ (indeed, it is skew-adjoint with respect to any $G$-invariant inner product on $E$). This implies that $a \cdot (a\cdot v) = 0$ if and only if $a \cdot v = 0$, from which the claim follows.
\end{proof}
We are now in a position to prove a result analogous to Proposition \ref{propm1} for this theory.
\begin{prop} \label{propym}
    Consider an extremal horizon in the theory \eqref{ymaction} with matter data $(\phi^I,A,B,C)$ on a compact cross-section $M$. Suppose the topological term satisfies the condition \eqref{ass2}. 
    \begin{enumerate}[(i)]
        \item If the horizon data is rotating, the Killing vector $K$ in Theorem~\ref{thm1} preserves $(\phi^I,A,B,C)$ in the sense of \eqref{kinv} with $\sigma_K = \iota_KA + \Gamma C$. \vspace{.1cm}
        \item The near-horizon matter fields \eqref{nhfields2} are preserved (in the sense of (\ref{kinv})) by the Killing vectors generating the isometries in Theorem~\ref{thm2} and, if the horizon data is both static and rotating, by the Killing vectors in Proposition~\ref{propads3}.
    \end{enumerate}
\end{prop}
\begin{proof}
Contracting (\ref{DFeom}) with $k$ and evaluating on $\mathcal{H}$, the condition (\ref{ass2}) shows that $k^\nu\mathcal{D}^\mu(h\mathcal{F}_{\mu\nu})$ vanishes on $\mathcal{H}$. Now
\begin{equation*}
    k^\nu\mathcal{D}^\mu(h\mathcal{F}_{\mu\nu}) = \mathcal{D}^\mu(hk^\nu\mathcal{F}_{\mu\nu}) - h\mathcal{F}_{\mu\nu}\mathcal{D}^\mu k^\nu \:\overset{\mathcal{H}}{=}\: -\mathcal{D}^\mu(hCk_\mu) = -h\mathcal{D}_kC,
\end{equation*}
where in the second step we used (\ref{Xdef}) and (\ref{tkk}) to show that $\mathcal{F}_{\mu\nu}\mathcal{D}^\mu k^\nu$ vanishes on $\mathcal{H}$.
Hence $\mathcal{D}_kC = 0$. Moreover, the identity $\mathcal{D}_k\mathcal{F} = 0$ pulled back to $M$ implies $\mathcal{D}_k B = 0$, and we have $\mathcal{D}_k\phi^I = 0$ as a consequence of (\ref{tkk}). Applying $\mathcal{D}_k$ to both equations in (\ref{fint}) and using the property (\ref{tkk}) of $\mathcal{F}$ to commute derivatives, we deduce $\mathcal{D}_k\psi^I = 0$ and $\mathcal{D}_k(\mathcal{D}_k H) = 0$. It now follows from Lemma \ref{lem} (taking $\Theta = \iota_Y H$ for any $Y$ tangent to $M$) that $\mathcal{D}_k H = 0$, so $H$ decouples from (\ref{fint}). To deal with $\psi^I$, introduce Ad-valued functions $\chi^I$ on $M$ by
\begin{equation*}
    \mathcal{D}\Phi^I \:\overset{\mathcal{H}}{=} D\phi^I +\chi^Ik.
\end{equation*}
Equivalently, $\chi^I = \mathcal{D}_r\Phi^I \vert_{r = 0}$ in Gaussian null coordinates. The identity $\mathcal{D}^2 \Phi^I = \mathcal{F} \cdot \Phi^I$ on $\mathcal{H}$ implies $\mathcal{D}_k \chi^I - \psi^I = C \cdot \phi^I.$
Applying $\mathcal{D}_k$ and Lemma \ref{lem} again, we obtain $\mathcal{D}_k \chi^I = 0$ and so $\psi^I = - C \cdot \phi^I$. Equation (\ref{fint}) reduces to 
\begin{equation} \label{fint2}
    D_K \phi^I = -\Gamma C \cdot \phi^I, \hspace{.8cm} \iota_K B = D(\Gamma C).
\end{equation}
Statement $(i)$ follows directly from (\ref{fint2}) (note that invariance of $B$ is implied by invariance of $A$):
\begin{align*}
    \Gamma\mathcal{L}_K C &= D_K(\Gamma C) - [\iota_KA,\Gamma C] = \Gamma[C, \iota_KA+\Gamma C],\\
    \mathcal{L}_K A &= \text{d}\iota_KA + \iota_K\text{d}A = D(\iota_KA) - [A,\iota_KA] + \iota_K B - \frac 12\iota_K [A,A] = D(\iota_KA + \Gamma C), \\
    \mathcal{L}_K\phi^I &= D_K\phi^I - \iota_KA\cdot\phi^I = -(\iota_KA + \Gamma C)\cdot\phi^I.
\end{align*}
The proof of $(ii)$ proceeds as in Proposition \ref{propm1}. In the gauge (\ref{nhfields2}) the near-horizon matter fields are invariant under $\partial_v$ and $v\partial_v - r\partial_r$. A computation as above shows that (\ref{kinv}) holds for the Killing vector $m$ in (\ref{mkv}) as well\footnote{In \cite{LL13} an additional global argument is used to establish the symmetry enhancement and constancy of (\ref{A}), which follows from the condition (\ref{fint2}) arising from the vanishing of $\gamma$.} with $\sigma_m = -v(\iota_K A + \Gamma C)$.  In the AdS$_3$ case we find from (\ref{fint2}) and the vanishing of (\ref{alphaex2}) that $C = 0$, and therefore also $\iota_K B = 0$. This implies the invariance of (\ref{nhfields2}) under any vector field tangent to the AdS$_3$ factor (i.e. in the span of $\partial_v,\partial_r$ and $K$) of the near-horizon geometry. In particular, this includes the $\mathfrak{so}(2,2)$ algebra of Killing fields in Proposition~\ref{propads3}.
\end{proof}
The fields $\Phi_\text{H}^I$ and $\mathcal{F}_{\text{H}}$ can be interpreted as the near-horizon limit of spacetime matter fields $\Phi^I,\mathcal{F}$ in Gaussian null coordinates in a gauge where $\mathcal{L}_k\Phi^I, \mathcal{L}_k\mathcal{A}$ and $\mathcal{L}_k \mathcal{F}$ all vanish (such a gauge exists as a consequence of (\ref{kinv})). However, as pointed out in \cite{LL13}, it is unclear whether this interpretation works for $\mathcal{A}_\text{H}$ because $\iota_k\mathcal{A}$ may be non-zero on $\mathcal{H}$ in this gauge. It follows from the proof of Proposition~\ref{propym} that all the matter data is annihilated by $\mathcal{D}_k$, so $\iota_k\mathcal{A}\vert_{\mathcal{H}}$ does not contribute to the horizon matter equations. In fact, it is possible to set $\iota_k\mathcal{A}\vert_{\mathcal{H}}$ to zero using the gauge transformation $\tau = \text{exp}(v\sigma)$ with $\sigma = \iota_k\mathcal{A} + rC$. For this we extend $C$ away from $M$ in any way such that $\mathcal{L}_kC = 0$ (note that $[\iota_k\mathcal{A},C] = 0$ on $\mathcal{H}$). Using (\ref{tkk}), the function $\sigma$ satisfies
\begin{equation*}
\mathcal{D}\sigma\:\overset{\mathcal{H}}{=}\:0, \hspace{.8cm} [\mathcal{F},\sigma] \:\overset{\mathcal{H}}{=}\:0,\hspace{.8cm}\sigma\cdot \Phi^I \:\overset{\mathcal{H}}{=}\: 0.
\end{equation*}
This implies that in the new gauge the leading order (in $r$) components of the spacetime matter fields are still annihilated by $\mathcal{L}_k$. Hence, although the near-horizon limit may be ill-defined, one can still think of (\ref{nhfields2}) as the leading order approximation to $(\Phi^I,\mathcal{A},\mathcal{F})$ away from $\mathcal{H}$ in this gauge.

The equations of motion induced by (\ref{DFeom}) on $M$ simplify due to (\ref{fint2}) and no longer contain the fields $\psi^I, \mathcal{D}_k\chi^I$ and $\mathcal{D}_kH$. In the absence of topological terms, they become
\begin{subequations}
\begin{align}
    &D^a(\Gamma hB_{ab}^i) - hK_bC^i - \tfrac18 \Gamma \left(f_{IJ}c^i_{AB}\overline{D_b\phi^{IA}}\phi^{JB} + \text{c.c.}\right) = 0, \\
    &\frac{1}{\Gamma}D_a(\Gamma f_{IJ}D^a\phi^{J}) -  f_{MN,IJ}\phi^J\langle D_a\phi^M,D^a\phi^N\rangle - 2 V_{,IJ}\phi^{J} + 2 h_{,IJ}\phi^{J}\text{Tr}\left(B_{ab}B^{ab} - 2C^2\right) = 0. \label{phiceom}
\end{align}
\end{subequations}
As observed in \cite{LL16} for a  complex scalar field, in many cases one can use the matter equations (\ref{phieom}, \ref{phiceom}) to show the fields $\phi^I$ must be trivial. In particular, if $h$ and $f_{IJ}$ are constant, $\mathcal{Q}_I$ vanishes and $V_{,IJ}$ is positive semi-definite, taking the inner product of (\ref{phiceom}) with $\phi^I$ (or of (\ref{phieom}) with $\phi^A$) and integrating over $M$ shows that $D\phi^I$ vanishes identically. This argument applies for example to the Einstein-Maxwell-charged Klein-Gordon model.

\section{Outlook}
In this work we proved symmetry enhancement results for extremal Killing horizons in theories with general matter content. We showed that any rotating extremal horizon admits a Killing field tangent to cross-sections and that any near-horizon geometry possesses at least a three-dimensional isometry group. These results require only the existence of a compact cross-section and the NEC. We demonstrated in various examples how the symmetries constrain the matter fields in the theory. The near-horizon isometry group is controlled by a constant $A$, which shares many properties with the surface gravity for non-extremal horizons. In particular, $A$ vanishes for doubly degenerate or triple horizons. In the context of the Aretakis instability for a scalar field, on such horizons there is an additional conserved quantity involving a second order transverse derivative of the field. 

Theorems \ref{thm1} and \ref{thm2} strongly constrain the geometry of extremal horizons and their associated near-horizon geometries. In four spacetime dimensions, the existence of the Killing vector $K$ reduces the horizon equations~(\ref{nhe}) to a system of ordinary differential equations. Within Einstein-Maxwell theory these can be solved explicitly, and any solution on $M =S^2$ is given by the extremal Kerr-Newman family \cite{LP02,KL09}. Hence the intrinsic rigidity theorem implies an analogue of the no-hair theorem for extremal horizons \cite{DL,CKL}.

In five dimensions the situation is considerably more complicated. Many near-horizon geometries are known \cite{KL13}, and the existence of a single Killing vector is no longer sufficient to solve the horizon equations. Known rotating solutions possess two commuting Killing vectors tangent to a cross-section. The question of whether a strengthening of Theorem \ref{thm1} guaranteeing the existence of a second Killing field can be established remains open. Even assuming the existence of two Killing fields, the results in this paper can be useful for constructing new solutions. As shown in Sections \ref{fsec} and \ref{gsec}, the matter equations simplify significantly as a consequence of the vanishing of the function $\gamma$ in (\ref{L}) or (\ref{gamma2}). We are currently investigating five-dimensional charged and rotating horizons using this formalism \cite{CL25}. Such horizons were recently studied numerically in \cite{HS24}.

The proof of Theorem \ref{thm1} relies heavily on a divergence identity (\ref{magid}), just like the corresponding results in \cite{DL,CKL,CD25}. Although equation (\ref{idint}) provides a direct derivation from the Einstein equations for the near-horizon geometry, it remains somewhat mysterious why the various terms in this identity can be arranged into total divergences and terms proportional to the divergence of the vector field $K$, and why the Killing vector takes the particular form (\ref{Kans}). It would be interesting to identify a geometric origin of this identity, which may involve the null vector $\ell$ in (\ref{ell}) that plays a crucial role in the argument.

The analysis in Section \ref{secare} suggests that for doubly degenerate horizons with surface gravity $\kappa$ and $A$ both equal to zero, there is a third constant $B$ controlling further degeneracy (i.e. vanishing for horizons on which the norm of the vector field (\ref{Vdef}) has a quadruple zero). Although we are not aware of any black hole spacetimes containing a horizon of multiplicity four or higher, it is possible that the pattern continues: for a horizon of multiplicity $n$ there might exist an ``$n$-th order surface gravity" which is constant and vanishes for multiplicity $n+1$. An interesting related question is whether $A$ or any of the higher order constants has a thermodynamical interpretation, analogous to the well-known relation between surface gravity and temperature.

Finally, one can ask about the link between such constants and the Aretakis instability. On a horizon of multiplicity $n > 2$ we can specify $\rho = \Gamma^{-1}r$ to $(n-1)$-th order in $r$ in such a way that $\partial_\rho^{l}(\nabla_iK^i) = 0$ on the horizon for $0 \leq l \leq n-2$. This requires solving PDEs of the form (\ref{pder}) for the first $n -2$ transverse derivatives of the function $\Gamma$. If for such a choice of $\rho$ the first $n-3$ derivatives of the function $A$ in (\ref{KAext}) vanish and $\partial_\rho^{n-2}A$ is constant on the horizon, Proposition \ref{areprop} can be generalised as follows: on a horizon of multiplicity $n > 1$, the quantity $I$ in (\ref{I}) and its first $n-2$ derivatives are conserved along the horizon, whereas $\partial_\rho^{n-1}I$ (which involves an $n$-th order $\rho$-derivative of the scalar field) generically grows linearly in the affine parameter $v$.

\appendix
\section{Extremal Kerr-Newman-de Sitter horizon } \label{Aa}
In this appendix we compute the quantities $A$ and $B$ considered in Sections \ref{enhsec} and \ref{ddhbigsec} for the extremal Kerr-Newman-de Sitter family. The metric depends on four parameters $(M,a,Q,l)$, where $l$ is related to the cosmological constant $\Lambda$ via $l^2 = 3\Lambda^{-1}$. In Boyer-Lindquist coordinates $(t,r,\theta,\phi)$, the metric reads
\begin{equation} \label{knds}
    \textbf{g} = -\frac{\Delta_r}{\Sigma}\left(\text{d}t - \frac{a\sin^2\theta}{\Xi}\text{d}\phi\right)^2 + \frac{\Delta_\theta\sin^2\theta}{\Sigma}\left(a\text{d}t - \frac{(r^2 + a^2)}{\Xi}\text{d}\phi\right)^2 + \Sigma\left(\frac{\text{d}r^2}{\Delta_r} + \frac{\text{d}\theta^2}{\Delta_\theta}\right).
\end{equation}
Here
\begin{align*}
    \Delta_r(r) &= (r^2 + a^2)\left(1 - \frac{r^2}{l^2}\right) - 2Mr + Q^2, \hspace{1cm} & \Xi &= 1 + \frac{a^2}{l^2}, \\
    \Delta_\theta(\theta) &= 1 + \frac{a^2}{l^2}\cos^2\theta, \hspace{1cm} & \Sigma(r,\theta) &= r^2 + a^2\cos^2\theta.
\end{align*}
The metric has Killing vectors $\xi = \partial_t$ and $m = \partial_\phi$. The horizons are located at the roots of $\Delta_r$. The spacetime contains an extremal horizon if two of the positive roots coincide, i.e. if there exists $r_0 > 0$ such that
\begin{equation*}
    \Delta_r(r_0) = 0, \hspace{1cm} \Delta_r'(r_0) = 0.
\end{equation*}
These conditions are most easily solved for $M$ and $Q$,
    \begin{align}
        M = r_0\left(1-\frac{a^2 + 2r_0^2}{l^2}\right), \hspace{1cm} Q^2 = r_0^2 - a^2-\frac{r_0^2(a^2 + 3r_0^2)}{l^2}.
    \end{align}
To obtain coordinates that are valid across $r = r_0$, we introduce ingoing coordinates $(v,r,\theta,\phi')$ by
\begin{equation*}
    \text{d}t = \text{d}v - \frac{(r^2 + a^2)}{\Delta_r}\text{d}r, \hspace{1cm} \text{d}\phi = \text{d}\phi' - \frac{a\Xi}{\Delta_r}\text{d}r.
\end{equation*}
If we further define  
\begin{equation*}
    \Omega_{\mathcal{H}} = \frac{a\Xi}{a^2 + r_0^2}, \hspace{1cm}\psi = \phi' - \Omega_{\mathcal{H}}v,
\end{equation*}
the generator $k = \xi + \Omega_{\mathcal{H}}m$ of the extremal horizon $\mathcal{H}$ at $r = r_0$ in coordinates $(v,r,\theta,\psi)$  simply becomes $k = \partial_v$. Note that we made a choice for the normalisation of $k$, which agrees with \cite{GLPP05}.

The function $\Gamma$ on $\mathcal{H}$ may be obtained by first calculating $X$ using (\ref{Xdef}),
\begin{equation}
    X = \frac{2a^2\cos \theta\sin\theta}{\Sigma(r_0,\theta)}\text{d}\theta + \frac{2ar_0(a^2 + r_0^2)\Delta_\theta(\theta)\sin^2\theta}{\Xi\Sigma(r_0,\theta)^2}\text{d}\psi.
\end{equation}
Since the Killing vector $K$ must be proportional to $m = \partial_\psi$ and the metric has no $\theta\psi$-components, $\Gamma$ must be such that $K^\flat = \Gamma X + \text{d}\Gamma$ has no d$\theta$ term. This leads to (compare \cite[Eq. 67]{KL09})
\begin{equation}
    \Gamma  = \frac{\Sigma(r_0,\theta)}{a^2 + r_0^2}= \frac{r_0^2 + a^2\cos^2\theta}{a^2 + r_0^2}.
\end{equation}
The normalisation of $\Gamma$ is chosen such that
\begin{equation} \label{normgam}
    k \:\overset{\mathcal{H}}{=}\:\Gamma \text{d}r.
\end{equation}
For this choice of $\Gamma$ the coordinate $r - r_0$ agrees with the coordinate $\rho$ introduced in Section \ref{enhsec} up to corrections of order $O(\rho^2)$. This also fixes the scaling freedom in the definition of $A$ and $B$. Observe that the extremal Kerr-Newman-dS spacetime satisfies the assumptions \eqref{a1}--\eqref{a2} in Section \ref{ddhsec} with $M = \{v=0,r=r_0\}$. Therefore, the constant $A$ may be calculated using the vector field
\begin{equation}
    V = k - \frac{\textbf{g}(k,m)}{\textbf{g}(m,m)}m.
\end{equation}
We have
\begin{equation}
    \textbf{g}(V,V) = -\frac{\Delta_\theta\Delta_r\Sigma}{(a^2 + r^2)^2\Delta_\theta-\Delta_ra^2\sin^2 \theta}.
\end{equation}
Using $r - r_0 = \rho + O(\rho^2)$ we obtain $\textbf{g}(V,V) = A\Gamma\rho^2 + O(\rho^3)$, where
\begin{equation}
    A = \frac{a^2 -l^2 + 6r_0^2}{l^2(a^2 + r_0^2)}.
\end{equation}
When the cosmological constant vanishes, this reduces to (with $M^2 = a^2 + Q^2$)
\begin{equation}
    A = -\frac{1}{Q^2 + 2a^2}.
\end{equation}
We see that $A$ is indeed a constant, which equals zero precisely when $a^2 = l^2 - 6r_0^2$. It may be verified that this is equivalent to $\Delta_r'' (r_0) = 0$, so that $A$ vanishes exactly when three horizons coincide. There is a two-parameter subfamily containing such a horizon, parametrised by $(r_0,l)$. For these spacetimes we may calculate $B$ as in (\ref{V3}),
\begin{equation}
B =  \frac{1}{6\Gamma}\partial_r^3\left[\textbf{g}(V,V)\right]\vert_{r = r_0} = \frac{4r_0}{l^2(l^2-5r_0^2)}.
\end{equation}
As claimed in Section \ref{secare}, this quantity is constant. It is non-vanishing, corresponding to the fact that the Kerr-Newman-dS family cannot contain a quadruple horizon (the fourth root of $\Delta_r$ necessarily has $r < 0$).

\section*{Statements and declarations}
\subsection*{Data availability} Data sharing is not applicable to this article as no datasets were generated or analysed during the current study.

\subsection*{Competing interests} The author has no relevant financial or non-financial interests to disclose.

\bibliographystyle{abbrv}
\bibliography{ref}

\end{document}